\documentclass[reprint,prx,twocolumn,superscriptaddress,longbibliography]{revtex4-1}
\usepackage{amsmath}
\usepackage{bbm}
\usepackage{graphicx}
\usepackage{amsfonts}
\usepackage{color}
\usepackage{upgreek}
\makeatletter
\def\captionof#1#2{{\def\@captype{#1}#2}}
\makeatother
\usepackage{hyperref}
\hypersetup{
    colorlinks = true,
    linkcolor =blue,
	citecolor=blue, 
	urlcolor=blue 
}
\begin{document}

\title{Emergent $\mathcal{PT}$ symmetry in a double-quantum-dot circuit QED set-up}

\author{Archak Purkayastha} 
\email{archak.p@tcd.ie}
\affiliation{Trinity College Dublin, The University of Dublin, College Green, Dublin, Ireland}

\author{Manas Kulkarni}
\email{manas.kulkarni@icts.res.in}
\affiliation{International Centre for Theoretical Sciences, Tata Institute of Fundamental Research,  Bangalore-560089, 
India}

\author{Yogesh N. Joglekar}
\email{yojoglek@iupui.edu}
\affiliation{Indiana University Purdue University Indianapolis (IUPUI), Indianapolis 46202, Indiana, USA}  
\date{\today}

\begin{abstract}
Open classical and quantum systems with effective parity-time ($\mathcal{PT}$) symmetry, over the past five years, have shown tremendous promise for advances in lasers, sensing, and non-reciprocal devices. And yet, how such effective $\mathcal{PT}$-symmetric non-Hermitian models emerge out of Hermitian quantum mechanics is not well understood. Here, starting from a fully Hermitian microscopic Hamiltonian description, we show that a non-Hermitian Hamiltonian emerges naturally in a double-quantum-dot-circuit-QED (DQD-circuit QED) set-up, which can be controllably tuned to the $\mathcal{PT}$-symmetric point. This effective Hamiltonian governs the dynamics of two coupled circuit-QED cavities with a voltage-biased DQD in one of them.  Our analysis also reveals the effect of quantum fluctuations on the $\mathcal{PT}$ symmetric system.  The $\mathcal{PT}$-transition is, then, observed both in the dynamics of cavity observables as well as via an input-output experiment.  As a simple application of the $\mathcal{PT}$-transition in this set-up, we show that loss-induced enhancement of amplification and lasing can be observed in the coupled cavities. By comparing our results with two conventional local Lindblad equations, we demonstrate the utility and limitations of the latter. Our results pave the way for an on-chip realization of a potentially scalable non-Hermitian system with a gain medium in quantum regime, as well as its potential applications for quantum technology.
\end{abstract}
\maketitle

\section{Introduction}
\label{Sec:intro}
For an isolated (quantum) system, the Hamiltonian is the generator of its time evolution. A fundamental postulate of the quantum theory is that this Hamiltonian is Hermitian. It ensures real energy eigenvalues, and a unitary time evolution for the system. 
This changed two decades ago, when Bender and co-workers discovered a large class of non-Hermitian, continuum Hamiltonians on an infinite line with purely real spectra~\cite{first_PT_theory}. The key feature shared by all of these Hamiltonians is an antilinear symmetry, i.e. they are invariant under the combined operations of parity ($\mathcal{P}$) and time reversal ($\mathcal{T}$). This antilinear symmetry, $[\mathcal{PT},\mathcal{H}]=0$, ensures that the eigenvalues of the Hamiltonian $\mathcal{H}$ are either purely real or occur in complex conjugate pairs~\cite{am2002,am2010}. When the eigenvalue $\lambda$ is real, the eigenstate $|\psi_\lambda\rangle$ is also a simultaneous eigenstate of the $\mathcal{PT}$ operator with eigenvalue one. When $\lambda$ is complex, the $\mathcal{PT}$ operator maps the eigenstate onto the eigenstate for its complex conjugate $\lambda^*$, i.e. $\mathcal{PT}|\psi_\lambda\rangle=|\psi_\lambda^*\rangle$. Therefore, a $\mathcal{PT}$-symmetric Hamiltonian spectrum shows a transition, known as the $\mathcal{PT}$-symmetry breaking transition, when it changes from purely real to complex. In addition to the eigenvalues, corresponding eigenstates also become degenerate at the $\mathcal{PT}$-transition point, and therefore the non-Hermitian Hamiltonian is defective at this exceptional point \cite{Kato1995,miri2019}. Over the years, it has become clear that $\mathcal{PT}$-symmetric Hamiltonians faithfully model open, classical, zero-temperature systems with balanced, spatially separated gain and loss; the $\mathcal{PT}$-symmetric phase with real eigenvalues thus represents an open, quasi-equilibrium system, whereas the $\mathcal{PT}$-broken phase with amplifying and decaying eigenvectors represents a system far removed from equilibrium.

The subject of non-Hermitian, $\mathcal{PT}$-symmetric Hamiltonians and their exceptional point degeneracies has evolved into a rich and active field. In the classical domain, effective non-Hermitian systems have been theoretically and experimentally studied in waveguides~\cite{Ruter2010,Encircling_EP_waveguide_expt_2018}, fiber loops~\cite{Regen2012}, resonators~\cite{Peng2014,Anti_PT_microcavity_expt_2020}, electrical circuits~\cite{RLC_PT,roberto2018,Topological_PT_circuit_expt_2020,
PT_breaking_circuit_expt_2019,EP_sensing_noise_theory_2019}, mechanical oscillators~\cite{BenderAJP2013,Elastodynamics_PT_expt_2020}, viscous fluids~\cite{Yu2014},  magnonics\cite{EP_surface_expt_2019,Non_hermitian_magnonics_expt_2019,
Magnon_anti_PT_2020_expt}, acoustics~\cite{Zhu2014,acoustic_PT_sensor,Acoustic_logic_gates_2020_expt}, optomechanics~\cite{Xu2016,Zhang2018}, optical lattices~\cite{Non_Hermitian_topology_expt_2019}. Occurrence of exceptional points has been used in device applications like sensing \cite{acoustic_PT_sensor,Non_Hermitian_Acoustics_topology_expt_2018,
Non_Hermitian_Acoustics_EP_expt_2018,Acoustic_logic_gates_2020_expt,
glucose_sensor_PT_2019}, single-mode lasing \cite{single_mode_laser_expt_2014}, unidirectional invisibility \cite{Unidirectional_invisibility_expt_2013}, loss-induced transparency \cite{Loss_induced_transparency_expt_2009}, loss-induced lasing \cite{loss_induced_lasing_expt}  etc (see recent reviews \cite{Ozdemir_2019_review,El-Ganainy_2018_review,Longhi_2017_review,Feng_2017_review,Konotop_2016_review} for further details and references).  A wide variety of physically motivated one-dimensional~(for example,~\cite{Longhi_topological_quasicrystals_2019,Non_Hermitian_MBL_2019,
Sushil_Mujumdar_non_Hermitian_2020,
Harter2018,Wunner2018,Harter2016,Liang2014,Joglekar2013,Joglekar2011,Vemuri2011}), and two-dimensional~(for example,~\cite{Non_Hermitian_dirac_theory_2020,Mock2016,Yoshida2019,Kawabata2019,
FoaTorres2019}) condensed-matter lattice models have also been studied. From a fundamental perspective, exploration of topological aspects of non-Hermitian systems has bee gaining ground both theoretically (for example,~\cite{Zhen2015,Xu2016,Harter2016,Harari2018,Gong2018,Kawabata2019,
FoaTorres2019,Longhi_topological_quasicrystals_2019}) and experimentally \cite{Bandres2018,Topological_PT_2020,
Topological_PT_circuit_expt_2020,Non_Hermitian_topology_expt_2019,Xiao2017}. In the quantum domain, realization of systems governed by effective non-Hermitian Hamiltonians and exploration of the effect of exceptional points has only been possible very recently.  Quantum non-Hermitian systems have been experimentally realized in linear optical circuits~\cite{Xiao2017,Passive_PT_critical_phenomena_expt_2019},  quantum photonics~\cite{Klauck2019}, ultracold atoms~\cite{Li2019,Non_hermitian_AB_ring_expt_2020}, a diamond NV center~\cite{Wu2019}, superconducting circuits~\cite{Naghiloo2019,EP_superconducting_circuits_expt_2019}, atom-light interacting systems \cite{Anti_PT_quantum_experiment_2020}, a lossy quantum-point-contact \cite{Non_hermitian_conductance_expt_2019}. 
\begin{figure*}
\includegraphics[width=\linewidth]{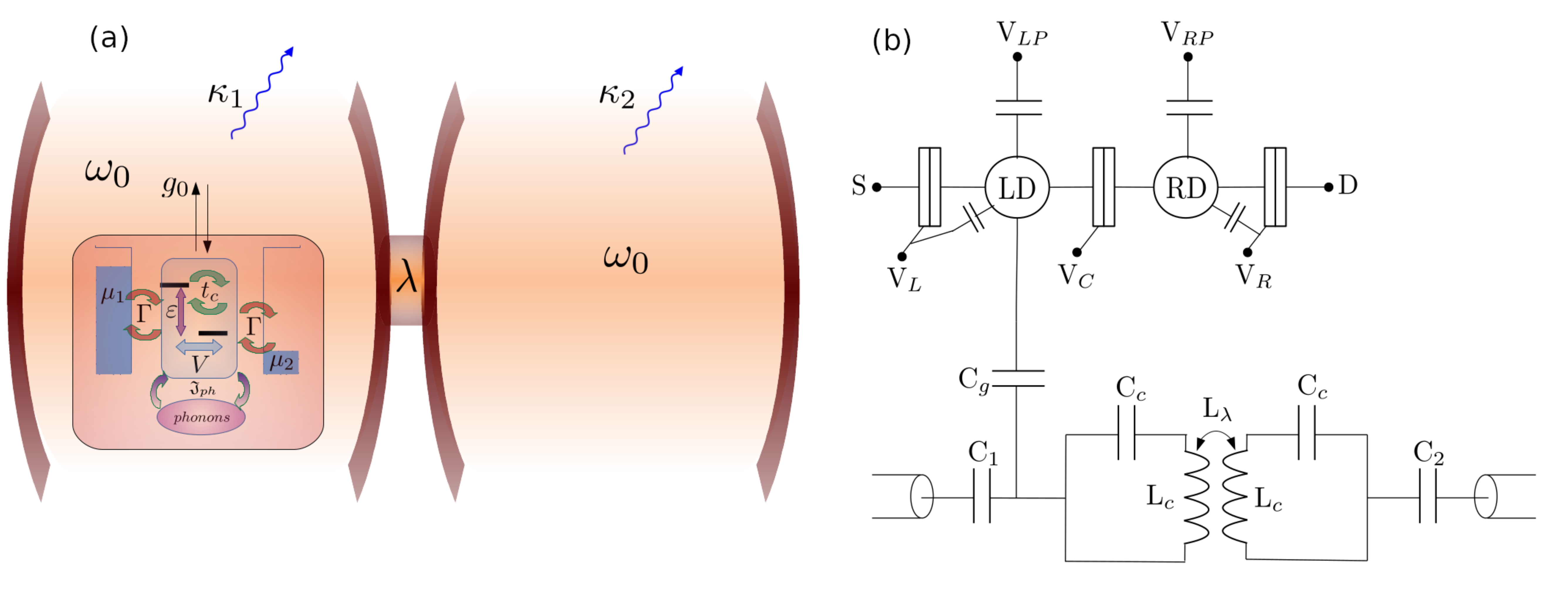}
\caption{ (color online) (a) Schematic diagram of the set-up we consider. (b) The equivalent circuit diagram. The set-up consists of two cavities each of frequency $\omega_0$ and losses $\kappa_{1,2}$, connected to each other with coupling $\lambda$. This is shown in (b) as two resonators of capacitance $C_c$ and inductance $L_c$, which are inductively coupled to each other via $L_\lambda$ and capacitively coupled to sources of loss via $C_1$ and $C_2$.  Additionally a DQD is located in the left cavity. The DQD is modelled by two fermionic sites (charge islands LD and RD  in the circuit diagram) with a detuning $\varepsilon$ between the on-site energies (controlled by the gate voltages $V_{LP}$, $V_{RP}$, $V_{L}$ and $V_{R}$), a hopping between the sites with strength $t_c$ (controlled by the gate-voltage $V_C$) and a strong repulsive interaction of strength $V$ between the sites. It is coupled to source (S) and drain (D) fermionic leads with coupling $\Gamma$ (controlled by $V_{LP}$, $V_{RP}$) via which a source-drain voltage bias of $\mu_1-\mu_2$ is applied across it. The DQD is coupled with the left cavity via dipole coupling of strength $g_0$ (shown in the circuit diagram by capacitive coupling $C_g$). To model experimental conditions accurately, we also consider the coupling of the DQD to the substrate phonons of spectral density $\mathfrak{J}_{ph}$. The voltage-biased DQD is configured to be population inverted, making it an effective gain medium \cite{photon_emission_DQD_cQED,DQD_micromaser,DQD_cQED_floquet_gain,
on_chip_DQD_light_source}. We show that, $\varepsilon$ and $t_c$ can be tuned so that the net gain in the left cavity is equal to the loss in the right cavity. Under this condition, the dynamics of the two cavities is governed by an effective $\mathcal{PT}$ symmetric Hamiltonian. We further explore signatures of exceptional points on tuning the coupling between cavities $\lambda$ and the loss at the right cavity $\kappa_2$. This set-up can be realized in state-of-the-art circuit QED platforms \cite{First_DQD_cQED_expt,Petta_review_2020,Clerk_Petta_Review_2020,
Kontos_DQD_cQED_review_2,wallraff2019_1}. In-situ tuning of coupling between cQED resonators, as well as of the resonator losses have been reported previously in experiments \cite{EP_superconducting_circuits_expt_2019,Tunable_loss_expt_2013,
Tunable_loss_expt_2014,Tunable_loss_expt_2016,
Tunable_coupling_expt_2015}.  } 
\label{fig:set_up}
\end{figure*}
However, all the realizations in the quantum regime are in cases where the overall system is dissipative. In particular, none of the realizations feature a gain medium. To our knowledge, transitions across exceptional points for balanced-gain-loss systems, featuring a gain medium in the quantum regime have not been realized yet, although there has been a theoretical proposal  \cite{PT_symmetric_cQED}.

The dynamics generated by a non-Hermitian Hamiltonian is not unitary, and such dynamics can result from an open quantum system coupled to one or more environments (baths). Non-Hermitian Hamiltonians arising out of phenomenological Lindblad equations has been theoretically explored in several recent works (for example, \cite{Avila2020,Quantum_PT_2020,Quantum_PT_2020_1,Quantum_PT_2020_3,
Yamamoto2019}). However, a complete microscopic theory of an open quantum system showing the emergence of a non-Hermitian Hamiltonian with a $\mathcal{PT}$-transition   starting from a fully Hermitian Hamiltonian description of a system coupled to multiple baths, has not yet been explored. In microscopic theories of open quantum systems, the full set-up of system+baths is taken to be an isolated system described by a Hermitian Hamiltonian. Then the evolution equation for the system degrees of freedom is obtained by integrating out the bath degrees of freedom. The resulting equation describes the non-unitary dynamics of the system. In the usual cases with a thermal bath, such an analysis leads to a dissipative (lossy) system, but, by clever bath-engineering, it can also lead to a gain~\cite{On_chip_laser_theory_DQD, superconducting_charge_qubit_lasing, DQD_resonator_lasing_theory, Kontos_2011, Kontos_DQD_cQED_review_2, DQD_cQED_laser_theory,
DQD_micromaser, DQD_cQED_floquet_gain, on_chip_DQD_light_source, Giant_photon_gain_DQD_cQED,photon_statistics_DQD_cQED}. However, due to the unitary nature of the underlying dynamics of the full set-up, each bath, in addition to providing loss or gain, generates a noise that is consistent with the fluctuation-dissipation theorem~\cite{Caves1982,Lau2018,Quantum_noise_EP_theory_2020,
EP_sensing_noise_theory_2019}. In the low-temperature limit, the noise generated by the lossy bath can be ignored depending on system parameters and time scales, but due to the quantum limits on linear amplifiers~\cite{Caves1982,Lau2018}, a gain-bath leads to noise down to zero temperature. Therefore, whether truly balanced-gain-loss $\mathcal{PT}$-symmetric quantum systems are possible, at the field-operator level~\cite{Scheel2018}, at the level of bosonic-field expectation values, or at the level of expectation values of bilinears of bosonic fields, is an open question. In particular, a fundamental analysis of a quantum system with gain and loss at low temperatures must take into account quantum fluctuations \cite{Kepesidis_2016}. 

In this work, starting from a completely microscopic Hermitian Hamiltonian description, we theoretically show that a state-of-the-art open quantum system can be tuned to observe the $\mathcal{PT}$-transition. The set-up we consider consists of a solid state double-quantum-dot (DQD) connected to two coupled circuit-QED (cQED) cavities. In recent years, a DQD in a cQED cavity has been well characterized both theoretically and experimentally \cite{Kontos_2011,Graphene_DQD_cQED_2015,On_chip_laser_theory_DQD,DQD_cQED_laser_theory,probing_electron_phonon_DQD_cQED_1,
probing_electron_phonon_DQD_cQED_2,Petta_review_2020,Clerk_Petta_Review_2020,
Kontos_DQD_cQED_review_2}. It is known that when the DQD is voltage-biased via electronic leads under suitable conditions, it can be population inverted and can act as a widely controllable gain medium \cite{DQD_cQED_laser_theory,photon_emission_DQD_cQED,
Giant_photon_gain_DQD_cQED,DQD_cQED_floquet_gain,photon_statistics_DQD_cQED}. This has led to the realization of an on-chip laser in the microwave regime~\cite{DQD_micromaser,on_chip_DQD_light_source}. 

In our analysis, the DQD is modeled by two fermionic sites, with hopping between them and strong repulsive interactions. The hopping strength and the on-site energies of the fermionic sites can be tuned widely in state-of-the-art realizations. Each fermionic site is also  connected to an external fermionic lead (bath), which allows us to apply a voltage bias across the DQD. Further, to model the DQD accurately, one should consider its coupling to the phonons in the substrate. We show here that a set-up of two coupled cavities with a DQD in one of them can be tuned so that time-evolution of the cavity operators is governed by an effective $\mathcal{PT}$-symmetric Hamiltonian. This happens when the effective gain from the DQD balances the cavity losses. However, the quantum fluctuations of the DQD leads to an additional noise term in the effective equations of motion for the two coupled cavities, which we derive from the microscopic model of the entire set-up. Thus, in this work, we propose an on-chip realization of a quantum $\mathcal{PT}$-symmetric system in which the effects of quantum fluctuations of the gain medium can be controllably studied.

Our proposal is very different from a recent proposal for the realization of a $\mathcal{PT}$-symmetric system in the cQED system by Quijandr{\'i}a et al.~\cite{PT_symmetric_cQED}. The latter considered two cQED cavities, each coupled to a qubit whose frequency is modulated via a coherent drive, and showed that the expectation values of the cavity field operators, in appropriate parameter regime, are governed by an effective $\mathcal{PT}$-symmetric Hamiltonian. The DQD voltage bias in our model acts as an incoherent drive, and our analysis extends to equations of motion for the cavity field operators and their bilinear combinations. This allows us to study the effect of quantum fluctuations on the cavities, which was not explored in Ref.~\cite{PT_symmetric_cQED}. As a simple application of the $\mathcal{PT}$-transition in our set-up, we further show the possibility of loss-induced enhancement of amplification and lasing. We further show loss-induced increase in average photon number in absence of any coherent drive in the cavities. This particular feature requires a gain medium with quantum fluctuations, and cannot be observed either in the classical systems with negligible fluctuations from the gain medium, or in the existing dissipative quantum system experiments without a gain medium discussed previously. Finally, we compare our results with more conventional local Lindblad approaches, highlighting universal features and the importance of the microscopic understanding, as well as pointing to the minimal Lindblad equation required to describe realistic quantum $\mathcal{PT}$-symmetric systems reasonably accurately in a parameter regime.

The paper is arranged as follows. In Sec.~\ref{Sec:set-up}, we describe the microscopic model of the set-up and give the parameter regime where it should be operated. In Sec.~\ref{Sec: Effective PT} we obtain the effective equations of motion for the field operators of the coupled cavities, and show how effective $\mathcal{PT}$-symmetry can be obtained. In Sec.~\ref{Sec: Balanced gain-loss} we explore the $\mathcal{PT}$-symmetric dynamics of the coupled cavities and the effect of quantum fluctuations when the cavity losses are balanced by the gain from the DQD. In Sec.~\ref{Sec: Dissipative PT} we show that effects of a $\mathcal{PT}$-transition can be observed even when the losses of the cavities are more than the gain from the DQD. We explore input-output experiments, effects of quantum fluctuations on them, as well as loss-induced lasing and enhancement of amplification in this section. In Sec.~\ref{Sec:local Lindblad} we compare our results with more conventional local Lindblad approaches. In Sec.~\ref{Sec: Conclusions} we summarize our main results and give the consequences of our work and future directions. Certain details of the analytical calculations are delegated to the Appendix.

\section{Microscopic Hamiltonian for the DQD coupled to circuit-QED cavities}
\label{Sec:set-up}

The schematic of the set-up we consider, along with the circuit diagram, is given in Fig.~\ref{fig:set_up}. It consists of two, identical coupled cavities, with a DQD in the left one.  The cavity with the DQD is an existing experimental set-up explored in a series of recent experiments \cite{wallraff2019_2,wallraff2019_1, photon_emission_DQD_cQED,DQD_micromaser,Phonon_assisted_gain_DQD_cQED,
DQD_cQED_floquet_gain,on_chip_DQD_light_source,
probing_electron_phonon_DQD_cQED_1,probing_electron_phonon_DQD_cQED_2,
Kontos_DQD_cQED_review_2}. The parameters of the DQD are widely tunable in experiment and, under a voltage bias, the DQD can be population inverted, thereby making it a tunable gain medium for the cavity. The main idea of this paper follows from the observation that when coupled to another cavity, the gain from the DQD can be tuned, within current state-of-the-art experimental parameters,  to exactly match the cavity losses, thereby realizing a $\mathcal{PT}$-symmetric system. 

We consider a completely microscopic model of the entire set-up.  The main parts of the set-up consist of the DQD and the two cavities. These are described by the following Hamiltonians
\begin{align}
&\hat{\mathcal{H}}_{DQD} = \frac{\varepsilon}{2}(\hat{c}_1^\dagger\hat{c}_1-\hat{c}_2^\dagger\hat{c}_2)+ t_c(\hat{c}_1^\dagger\hat{c}_2 + \hat{c}_2^\dagger\hat{c}_1)+V\hat{c}_1^\dagger\hat{c}_1\hat{c}_2^\dagger\hat{c}_2 , \nonumber \\
& \hat{\mathcal{H}}_{C}=\omega_{0}(\hat{b}^{\dagger}_1\hat{b}_1 +\hat{b}^{\dagger}_2 \hat{b}_2) + \lambda ( \hat{b}^{\dagger}_1 \hat{b}_2 + \hat{b}^{\dagger}_2 \hat{b}_1), \nonumber \\
&\hat{\mathcal{H}}_{DQD-C} =  g_0  \Theta(t) (\hat{c}_1^\dagger\hat{c}_1-\hat{c}_2^\dagger\hat{c}_2)(\hat{b}_1^\dagger + \hat{b}_1),
\label{eq:DQDC}
\end{align}
where $\Theta(t)$ is the Heaviside function. Here, $\hat{c}^\dagger_{1,2}$ denote fermionic creation operators for sites 1 and 2 that model the DQD, $\varepsilon$ is the energy difference between the two sites, $t_c$ is the hopping amplitude between the two sites, and $V>0$ denotes the capacitive charging energy between the two sites. $\hat{b}_{1,2}$ represent the bosonic creation operators for the two cavities, each with frequency $\omega_0$, that are coupled via a number-conserving hopping process with amplitude $\lambda$. Closely following the experimental set-ups, the DQD is dipole-coupled to the (first) cavity with strength $g_0$ when the cavity is switched on at time $t=0$. Thus, Eq.(\ref{eq:DQDC}) captures the microscopic model of the DQD and the two cavities. 

Each of the three main components (two cavities and one DQD) is connected to multiple baths. The two bosonic baths, one for each cavity, are described by the Hamiltonian
\begin{align}
\hat{\mathcal{H}}_{B} = \sum_{s=1}^{\infty}(\Omega_{s1} \hat{B}_{s1}^\dagger \hat{B}_{s1}+\Omega_{s2} \hat{B}_{s2}^\dagger \hat{B}_{s2}),
\end{align}
where $\hat{B}^\dagger_{s1}$ ($\hat{B}^\dagger_{s2}$) is the bosonic creation operator of the $s$th mode of the bath attached to the first (second) cavity, and $\Omega_{s1}$ ($\Omega_{s2}$) is the energy that mode. The baths are coupled to the respective cavities as  
\begin{align}
\hat{\mathcal{H}}_{C-B}= \sum_{\ell=1}^2 \sum_{s=1}^\infty \kappa_{s\ell} (\hat{B}_{s\ell}^\dagger \hat{b}_\ell + \hat{b}_\ell^\dagger \hat{B}_{s\ell}),
\end{align}
where $\kappa_{s\ell}$ is the coupling of the $s$th mode of the bath attached to the $\ell$th cavity. Each fermionic site of the DQD  is connected to its respective lead, are described by the Hamiltonian 
\begin{align}
&\hat{\mathcal{H}}_{L}=\sum_{s=1}^{\infty}\left(\mathcal{E}_{s1} \hat{a}_{s1}^\dagger \hat{a}_{s1}+\mathcal{E}_{s2} \hat{a}_{s2}^\dagger \hat{a}_{s2}\right), 
\end{align}
where $\hat{a}^\dagger_{s1}$ ($\hat{a}^\dagger_{s2}$) is the fermionic creation operator of the $s$th mode of the source (drain)
lead attached to the first (second) dot, and $\mathcal{E}_{s1}$ ($\mathcal{E}_{s2}$) is the energy of that mode. The fermionic leads are bilinearly coupled to the DQD via 
\begin{align}
&\hat{\mathcal{H}}_{DQD-L}=\sum_{\ell=1}^2 \sum_{s=1}^\infty \Gamma_{s\ell} (\hat{a}_{s\ell}^\dagger \hat{c}_\ell + \hat{c}_\ell^\dagger \hat{a}_{s\ell}).
\end{align}
where $\Gamma_{s\ell}$ is its coupling of the $s$th mode of the fermionic lead attached to the DQD site $\ell$. To model experimental set-up consistently, we also have to consider that the DQD is dipole-coupled to a phononic bath in the substrate on which it is located \cite{probing_electron_phonon_DQD_cQED_1}. The phonon bath and the coupling Hamiltonians are given by 
\begin{align}
&\hat{\mathcal{H}}_{ph} = \sum_{s=1}^{\infty}\Omega_{s}^{ph} \hat{B}_{s}^{ph\dagger} \hat{B}_{s}^{ph},  \\
&\hat{\mathcal{H}}_{DQD-ph} = \left( \hat{c}_1^\dagger\hat{c}_1-\hat{c}_2^\dagger\hat{c}_2\right) \sum_{s=1}^\infty \lambda_s^{ph} \left(\hat{B}_{s}^{ph\dagger} + \hat{B}_{s}^{ph}\right). \nonumber
\end{align}
where $\hat{B}_{s}^{ph}$ is the phonon annihilation operator for the $s$th mode of the phononic bath, and $\lambda^{ph}_s$ is its coupling to the DQD dipole operator. While this piece of the microscopic Hamiltonian is not necessary for the physics that we will discuss, it is unavoidable, and relevant, in some of the state-of-the-art experimental set-ups \cite{photon_emission_DQD_cQED,DQD_micromaser,Phonon_assisted_gain_DQD_cQED,
DQD_cQED_floquet_gain,on_chip_DQD_light_source,
probing_electron_phonon_DQD_cQED_1,probing_electron_phonon_DQD_cQED_2}. 

The two bosonic baths for the two cavities, the two fermionic leads, and the phonon bath are characterized by their frequency-dependent spectral functions, 
\begin{align}
\label{all_spectral_functions}
&\mathfrak{J}_\ell(\omega) = \sum_{s=1}^\infty |\kappa_{s\ell}|^2 \delta(\omega - \Omega_{s\ell})\simeq\kappa_\ell,~~\forall~\omega\gg 0 \nonumber \\
&\mathfrak{J}^f_{\ell}(\omega) = \sum_{s=1}^\infty|\Gamma_{s\ell}|^2 \delta(\omega - \mathcal{E}_{s\ell})=\Gamma, \nonumber \\
& \mathfrak{J}_{ph}(\omega) = \sum_{s=1}^\infty |\lambda_{s}^{ph}|^2 \delta(\omega - \Omega_{s}^{ph}) \nonumber \\
&=\gamma_b\omega\left[1-\textrm{sinc}(\omega/\omega_{c})\right]e^{-{\omega^2}/{2\omega_{\max}^2}}.
\end{align}
Here $\mathfrak{J}_\ell(\omega)$ is the spectral function of the bosonic bath coupled to $\ell$th cavity, $\mathfrak{J}^f_{\ell}(\omega)$ is the spectral function of the fermionic lead coupled to $\ell$th site in the DQD, and $\mathfrak{J}_{ph}(\omega)$ is the spectral function of the phononic bath. For simplicity, we consider both fermionic leads to be in the ``wide-bath limit''; that gives rise to a constant spectral function denoted by $\Gamma$. The spectral functions for the bosonic baths of the two cavities give rise to individual decay rates $\kappa_\ell$. The effect of the substrate phonons on the DQD, as well as the spectral function for the phonons can vary depending on the platform. They have been well-characterized in literature both theoretically \cite{phonon_spectral_deriv,DQD_SSC,
probing_electron_phonon_DQD_cQED_1} and experimentally \cite{measuring_phonons_2010,probing_electron_phonon_DQD_cQED_2}. The spectral function $\mathfrak{J}_{ph}(\omega)$ in Eq.(\ref{all_spectral_functions}) is known to be a good description for gate-defined DQDs on a GaAs substrate \cite{
probing_electron_phonon_DQD_cQED_1,phonon_spectral_deriv}. The frequency $\omega_c$ is given by $\omega_c=c_n/d$, where $c_n$ is the speed of sound in GaAs and $d$ is the distance between the two quantum dots. The frequency $\omega_{\max}$ is the upper cut-off that provides spectral damping at frequencies much higher than the repetition rate of phonon travel between the two dots. We take $\omega_{\max}=10\omega_c$. For a DQD with GaAs substrate, $c_n\approx 3\times10^4$ m/s and $d\approx 150$ nm, gives $\omega_c=20$GHz. The dimensionless parameter $\gamma_b$ controls the coupling between the DQD and the phonons. We will like to mention that, although not required, microscopic models giving rise to all the spectral functions given in Eq.(\ref{all_spectral_functions}) can be written down assuming finite but large hard-cutoffs in frequency \cite{Nazir2018}.

The initial state of the whole set-up is the direct product of arbitrary states of the DQD and the cavities, and equilibrium thermal states of the baths with their respective temperatures and chemical potentials. We consider all the baths at the same inverse temperature $\beta$, but the chemical potentials for the two fermionic leads are given by $\mu_1\neq\mu_2$ (see Appendix~\ref{Derivation}). This creates a voltage bias across the DQD, thereby driving the DQD to a non-equilibrium steady state. We consider that the connection between the DQD, the fermionic leads and the phononic substrate is switched on at a time $t=t_0 \ll 0$. On the other hand, the two cavities are connected to the DQD and the bosonic baths at time $t=0$. Further, as we discuss below, we make assumptions on the various energy scales appearing in the full set-up.

The DQD Hamiltonian is diagonalized by rotating to the a new basis 
\begin{align}
\label{def_theta}
&\hat{A}_\ell=\sum_{\ell'}R_y^\dagger(\theta)_{\ell\ell'}\hat{c}_{\ell'}, ~
R_y(\theta)=\exp(-i\theta\sigma_y/2), \nonumber \\
&\theta=\arctan(2t_c/\varepsilon),
\end{align}
where  $\sigma_y$ is the Pauli y-matrix. In the rotated basis, the DQD Hamiltonian becomes 
\begin{align}
&\hat{\mathcal{H}}_{DQD} =\frac{\omega_q}{2}(\hat{N}_1 - \hat{N}_2) + V\hat{N}_1\hat{N}_2, 
\end{align}
where $\omega_q=\sqrt{\varepsilon^2+4t_c^2}$,  and $\hat{N}_\ell=\hat{A}_\ell^\dagger \hat{A}_\ell$ ($\ell=1,2$). We look at the parameter regime that ensures a resonant DQD at low temperature, 
\begin{align}
\label{conditions_energy_scales_temp}
\omega_0=\omega_q,~~\beta\omega_0\gg 1,
\end{align} 
a weak cavity-DQD couplng, i.e. 
\begin{align}
\label{conditions_energy_scales0}
g_0\sqrt{n_\mathrm{photons}}\lesssim \Gamma, 
\end{align}
where $n_\mathrm{photons}$ is the average number of photons in the cavity coupled to the DQD, and 
\begin{align}
\label{conditions_energy_scales}
&\omega_0/2\ll -\mu_2,\mu_1\ll V, \nonumber\\
&\kappa_1,\kappa_2\ll \lambda \ll \Gamma\ll\omega_0.
\end{align}
 In experiments, the widely controllable parameters of the DQD are $\varepsilon$ and $t_c$, which thereby allow us to tune $\omega_q$ and $\theta$. Thus, for a $\omega_0$, the resonant condition $\omega_0=\omega_q$ can be satisfied by tuning $\varepsilon$ and $t_c$. Under this condition, since the DQD-cavity coupling is weak, we can simplify the dipole coupling Hamiltonian through rotating wave approximation to
\begin{align}
&\hat{\mathcal{H}}_{DQD-C} = -g\Theta(t) (\hat{A}_2^\dagger\hat{A}_1 \hat{b}_1^\dagger+\hat{A}_1^\dagger\hat{A}_2 \hat{b}_1)
\end{align}
where the effective coupling is given by 
\begin{align}
\label{def_g}
g=g_0(2t_c/\omega_0)=g_0\sin\theta.
\end{align}
It is possible to tune $\theta$ while maintaining the DQD-cavity resonance condition $\omega_0=\omega_q$ by varying $\varepsilon$ and $t_c$ along the ellipse defined by $\omega_0^2=\varepsilon^2+4t_c^2$. As we will see below, this freedom allows us to tune the system such that the effective dynamics of the two cavities is described by a $\mathcal{PT}$-symmetric Hamiltonian.

\section{Effective Hamiltonian and emergent $\mathcal{PT}$ symmetry}
\label{Sec: Effective PT}
We can obtain effective equations of motion for the bosonic cavity operators by integrating out the rest of the set-up (see Appendix~\ref{Derivation}). When we are in the appropriate parameter regime (see Eqs.(\ref{conditions_energy_scales0}), (\ref{conditions_energy_scales})), this leads to
\begin{align}
\label{eq:eom}
&i\frac{d}{dt}\left(
\begin{array}{c}
\hat{b}_1\\
\hat{b}_2\\
\end{array}
\right) = \mathbf{H}_{\mathrm{eff}}\left(
\begin{array}{c}
\hat{b}_1\\
\hat{b}_2\\
\end{array}
\right) +\left(
\begin{array}{c}
\hat{\xi}_A\\
0\\
\end{array}
\right)- E_0e^{-i\omega_d t}\left(
\begin{array}{c}
0\\
1\\
\end{array}
\right), 
\end{align}
where, for future reference, we have additionally added a coherent drive of strength $E_0$ and frequency $\omega_d$ in the cavity without the DQD.
The 2$\times$2 non-Hermitian Hamiltonian is given by 
\begin{align}
\label{def_H_eff}
&\mathbf{H}_{\mathrm{eff}}=\left[
\begin{array}{cc}
\omega_0-i\frac{\kappa_1}{2}+i\Gamma\delta & \lambda\left(1-\delta\right)\\
\lambda & \omega_0-i\frac{\kappa_2}{2}\\
\end{array}\right],
\end{align} 
where 
\begin{align}
\label{def_delta}
\delta=g^2\Delta N_{ss}/\Gamma^2,
\end{align}
$\delta< 1$ by the choice of our parameters and $\Delta N_{ss}=\langle \hat{N}_1 \rangle_{ss}-\langle \hat{N}_2 \rangle_{ss},$ is the steady-state population inversion in the DQD in the absence of cavities. The operator $\hat{\xi}_A(t)$ is embodies the noise resulting from the presence of the DQD, in accordance with fluctuation-dissipation theorem. It has a Lorenzian power spectrum,
\begin{align}
\label{eq:noisevariance}
& \langle\hat{\xi}_A^\dagger(t)\hat{\xi}_A(t^\prime)\rangle\simeq \int_{-\infty}^\infty \frac{d\omega}{2\pi} P(\omega)   e^{i\omega(t-t^\prime)}, \\
& P(\omega)=g^2\langle \hat{N}_1 \rangle_{ss} \frac{2\Gamma}{(\omega-\omega_0)^2+\Gamma^2} \nonumber,
\end{align}
and a mean value, 
$\langle\hat{\xi}_A(t)\rangle=g\langle \hat{A}_2^\dagger \hat{A}_1 \rangle_{ss}$, 
where $\langle \hat{A}_2^\dagger \hat{A}_1 \rangle_{ss}$ is the steady state coherence of the DQD in absence of the two cavities. Note that the properties of the noise operator are not phenomenological, but are microscopically derived from our model. The noise is not delta-correlated in time, as sometimes assumed in semi-classical approaches (for example, in Ref.\cite{Kepesidis_2016}). Under our assumptions on energy scales (Eqs.(\ref{conditions_energy_scales0}),(\ref{conditions_energy_scales})) $\langle\hat{A}_2^\dagger \hat{A}_1 \rangle_{ss}\propto \Gamma$, so that $g\langle \hat{A}_2^\dagger \hat{A}_1 \rangle_{ss}/\omega_0\lesssim \left(\Gamma/\omega_0\right)^2\ll 1$. In this condition, it can be checked from Eq.(\ref{def_H_eff}) that $\langle\hat{\xi}_A(t)\rangle$ has a negligible effect on the dynamics of the cavities as long as  
\begin{align}
\label{approx2}
\sqrt{n_{photons}}\gg g\langle \hat{A}_2^\dagger \hat{A}_1 \rangle_{ss}/\omega_0.
\end{align}
So, assuming this, we can set
\begin{align}
\label{approx3}
\langle\hat{\xi}_A(t)\rangle\approx 0.
\end{align}
In all of the equations above, the expectation value represents quantum statistical average, i.e. $\langle O\rangle=\mathrm{Tr}(\rho_{\mathrm{tot}}O)$ where $\rho_{\mathrm{tot}}$ is the density matrix for the initial state of the whole set-up (see Appendix~\ref{Derivation}). We have also numerically checked that Eq.(\ref{approx3}) holds in our chosen parameter regime. On the other hand, as we will see, $\langle\hat{\xi}_A^\dagger(t)\hat{\xi}_A(t^\prime)\rangle$, which embodies the quantum fluctuations due to the DQD, cannot be neglected.  The bosonic baths that provide dissipation for the two cavities also lead to thermal noise. However, it can be shown that at low temperatures, i.e, for $\beta \omega_0\gg 1$, their contribution is negligible compared the fluctuations coming from the DQD (see Appendix~\ref{thermal_fluctuations}). So, we have ignored them in Eq.(\ref{eq:eom}).  

We note that $\mathbf{H}_{\mathrm{eff}}$ is non-Hermitian in two ways. First, its diagonal elements have imaginary parts that represent the coupling-to-the-bosonic bath losses $\kappa_\ell$ and the gain $\Gamma\delta$ for the first cavity which houses the DQD. Secondly, the population inversion in the DQD has suppressed the hopping from the second cavity to the first one, leading to real, asymmetric off-diagonal elements. The presence of two non-Hermiticities makes $\mathbf{H}_{\mathrm{eff}}$ different from the most commonly studied 2$\times$2 $\mathcal{PT}$-symmetric Hamiltonians. It is straightforward to obtain the quadratic characteristic equation for $\mathbf{H}_\mathrm{eff}$ and check that it is purely real if and only if 
\begin{align}
\label{eq:realchar}
\kappa_1+\kappa_2=2\Gamma\delta=2(g_0\sin\theta)^2\Delta N_{ss}/\Gamma.\\\textrm{(balanced gain-loss)}\nonumber
\end{align}
Eq.(\ref{eq:realchar}) ensures that the eigenvalues of $\mathbf{H}_{\mathrm{eff}}$ are either purely real or occur in complex conjugate pairs, and thus endows $\mathbf{H}_{\mathrm{eff}}$ with some antilinear symmetry. With this constraint, the eigenvalues $\Lambda_\pm$ of the effective, non-Hermitian Hamiltonian, Eq.(\ref{def_H_eff}) are given by
\begin{align}
\label{Lamda_pm_balanced_gain_loss}
\Lambda_{\pm}~=~\omega_0~\pm~\sqrt{\lambda^2(1-\delta)-\left(\frac{\kappa_2}{2}\right)^2}\\~\textrm{(for balanced gain-loss)}, \nonumber
\end{align}
 with an exceptional point degeneracy occuring at
\begin{align}
\label{lamda_EP_balanced_gain_loss}
\lambda_{EP}=\kappa_2/(2\sqrt{1-\delta})~~~\textrm{(for balanced gain-loss)}
\end{align}
 When the gain and loss cavities are strongly coupled, $\lambda\geq\lambda_{EP}$, the system is in the $\mathcal{PT}$-symmetric phase (real spectrum); a weak gain-loss coupling, $\lambda<\lambda_{EP}$ drives the system into the $\mathcal{PT}$-broken region (complex conjugate spectrum). These properties are reminiscent of the standard $\mathcal{PT}$-transition; however, $\mathbf{H}_{\mathrm{eff}}$ is not parity-time symmetric when parity exchanges the cavity labels $1\leftrightarrow 2$, and time-reversal is just the complex conjugation operation $\mathcal{K}$. To obtain the corresponding antilinear $\mathcal{PT}$ operator for this model, we express $\mathbf{H}_\mathrm{eff}$ as a linear combination of identity and Pauli matrix components,
\begin{align}
\label{eq:Heff}
&\mathbf{H}_{\mathrm{eff}}=\omega_0\mathbbm{1}_2+\frac{\lambda}{2}(2-\delta)\sigma_x-i\frac{\lambda\delta}{2}\sigma_y+i\frac{\kappa_2}{2}\sigma_z. 
\end{align}
It follows from Eq.(\ref{eq:Heff}) that the Hamiltonian can be cast into the traditional form with a rotation $R_x(\phi)=\exp(-i\phi\sigma_x/2)$ where $\phi=\arctan(\lambda\delta/2\kappa_2)$. In the transformed basis, the Hamiltonian $H'=R_x^\dagger(\phi)\mathbf{H}_{\mathrm{eff}}R_x(\phi)$ becomes 
\begin{align}
\label{eq:Hprime}
&H'=\omega_0\mathbbm{1}_2+\lambda(1-\delta/2)\sigma_x+i\frac{\kappa_2}{2}\sec(\phi)\sigma'_z.
\end{align}
The traditional Hamiltonian $H'$ is symmetric under interchange of cavity labels (parity) and complex conjugation operation $\mathcal{K}$ (time-reversal). Due to the antilinear nature of the latter, though, in the original basis, the $\mathcal{PT}$ operator becomes to 
\begin{align}
\label{eq:pt}
&\mathcal{PT}=R_x(\phi)\left[\sigma_x\mathcal{K}\right]R_x^\dagger(\phi)=R_x(2\phi)\sigma_x\mathcal{K}.
\end{align}
It is straightforward to check that the antilinear operator in Eq.(\ref{eq:pt}) commutes with the non-Hermitian Hamiltonian $\mathbf{H}_{\mathrm{eff}}$. Thus, the DQD circuit-QED set up provides a non-trivial example of the antilinear symmetry that arises from the steady-state population inversion, $\delta=2g^2\Delta N_{ss}/\Gamma^2\neq 0$. 
\begin{table}[!t]
\begin{tabular}{| c | c |}\hline
$\omega_0$ & 8 GHz \\
$g_0$ & 60 MHz \\
$\Gamma$ & 90 MHz \\
$\mu_1=-\mu_2$ & 30 GHz \\ 
$\beta$ & 10 GHz$^{-1}$ \\ \hline
\end{tabular}
\caption{The above table gives our chosen parameters for numerical calculations.  These values fall within the window of parameters in state-of-the-art experiments on DQD-cQED systems \cite{Phonon_assisted_gain_DQD_cQED,
DQD_cQED_floquet_gain,
probing_electron_phonon_DQD_cQED_1,probing_electron_phonon_DQD_cQED_2,
Kontos_DQD_cQED_review_2,wallraff2019_1}.}
\label{Table:parameters}
\end{table}

\begin{figure}[!h]
\begin{tabular}{c}  
\includegraphics[width=\columnwidth]{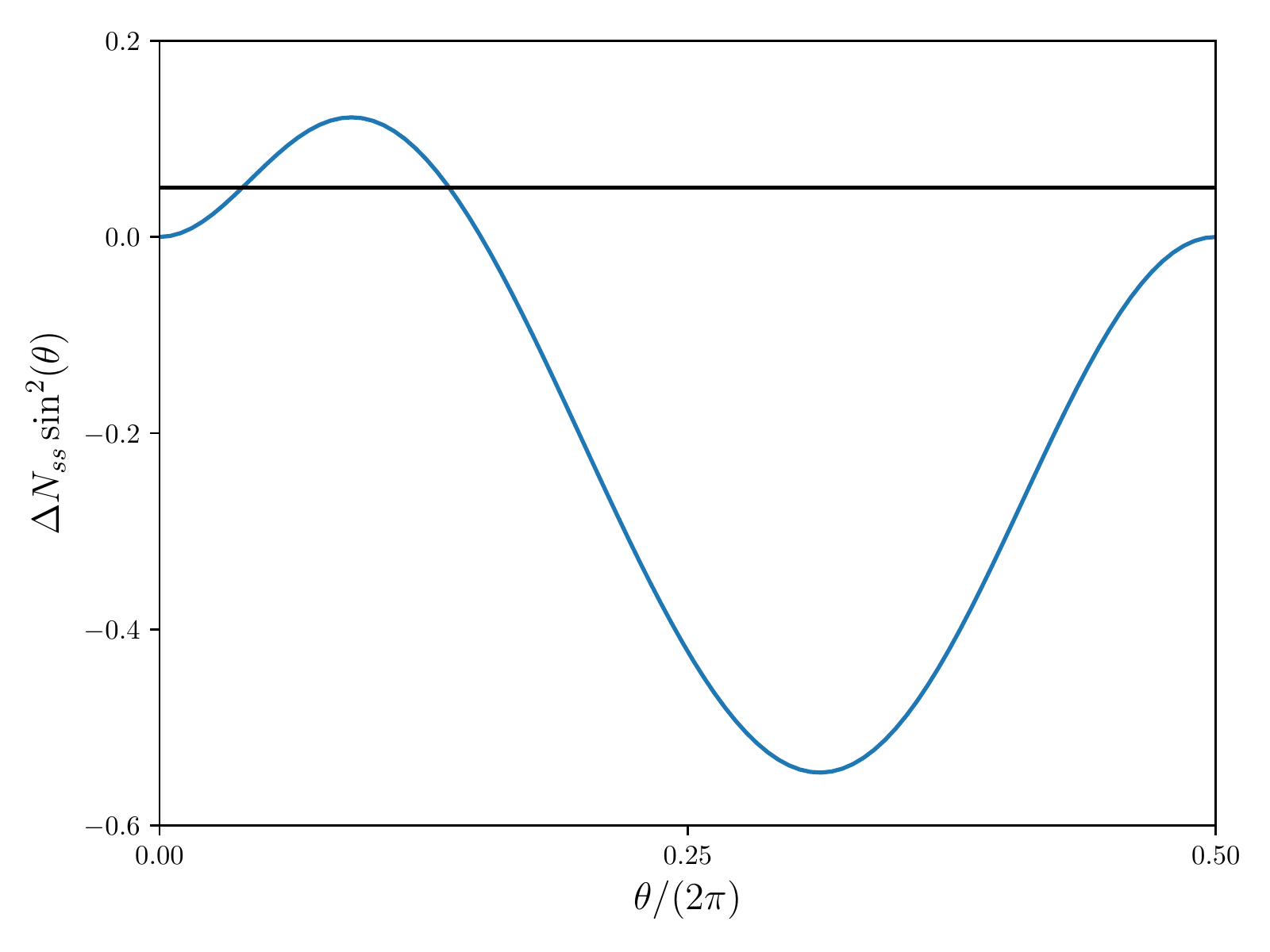} \\
$\mathcal{PT}$-symmetric points:\\
\begin{tabular}{| c | c |}\hline
$\varepsilon$ (GHz) & $t_c$ (GHz) \\ \hline
$7.760$ & $ 0.973$ \\
$5.208$ & $ 3.036$ \\ \hline
\end{tabular}
\end{tabular}
\caption{(color online) Numerically obtained $\Delta N_{ss}\sin^2\theta$ as a function of $\theta$ obtained by varying $\varepsilon$ and $t_c$ while maintaining the resonance-condition for the DQD, i.e. $\omega_q=\omega_0$. Here, we have chosen $\kappa_1=\kappa_2= 2 MHz$. The other system parameters are given in Table~\ref{Table:parameters}. The horizontal line shows the right-hand side of Eq.(\ref{eq:nss}). The dynamics of the coupled cavities is governed by an effective $\mathcal{PT}$-symmetric Hamiltonian at the intersection points of the curve and the line. Corresponding values of $\varepsilon$ and $t_c$ are listed in the table above. These values fall within accessible parameter values in state-of-the-art DQD-cQED experiments \cite{Phonon_assisted_gain_DQD_cQED,
DQD_cQED_floquet_gain,
probing_electron_phonon_DQD_cQED_1,probing_electron_phonon_DQD_cQED_2,
Kontos_DQD_cQED_review_2,wallraff2019_1}. }
\label{fig:N_ss_sin}
\end{figure}
Having established the antilinear symmetry of the non-Hermitian Hamiltonian $\mathbf{H}_\mathrm{eff}$ under the constraint of Eq.(\ref{eq:realchar}), we now show that the constraint can be satisfied. Since the steady-state population inversion $\Delta N_{ss}(\theta)$ depends on the DQD parameters, we recast Eq.(\ref{eq:realchar}) as 
\begin{align}
\label{eq:nss}
&\Delta N_{ss}\sin^2\theta=\frac{\Gamma(\kappa_1+\kappa_2)}{2g_0^2},
\end{align}
where the $\theta$-dependent quantities are only present on the left-hand side of the equation. We consider a realistic set of parameters for the system (Table~\ref{Table:parameters}), and plot both sides of Eq.(\ref{eq:nss}) as a function of $\theta$. 
Figure~\ref{fig:N_ss_sin} shows the $\theta$-dependence of two sides of Eq.(\ref{eq:nss}), with the flat line indicating the right-hand side. The DQD on-site energy $\varepsilon$ and hopping amplitude $t_c$ necessary to satisfy Eq.(\ref{eq:nss}) and make the set-up $\mathcal{PT}$ symmetric are shown in the same figure. These values are within the reach of state-of-the-art experiments. For a given value of $\kappa_2$, there can be two values of $\varepsilon$ and $t_c$ which satisfy the balanced-gain-loss condition. It is interesting to note that these two conditions can have widely different values of population inversion. For our choice of parameters, we have
\begin{align}
\label{population_inversion}
\Delta N_{ss} = 0.846~~{\rm~for~}\varepsilon=7.760{\rm~GHz},~t_c=0.973{\rm~ GHz} \nonumber \\
\Delta N_{ss} = 0.087~~{\rm~for~}\varepsilon=5.208{\rm~GHz},~t_c=3.036{\rm ~GHz}.
\end{align} 
In particular, the second condition has a very low population inversion. Because of the $\sin^2\theta$ factor, both these values of population inversion give the same value of $\delta$, thereby leading to the same $\mathbf{H}_\mathrm{eff}$. However, the strength of quantum fluctuations of the gain medium depends on $g^2\langle \hat{N}_1 \rangle_{ss}$ (see Eq.(\ref{eq:noisevariance})), and will be different in the two cases.

If Eq.(\ref{eq:realchar}) is not satisfied, $\mathbf{H}_\mathrm{eff}$ is no longer $\mathcal{PT}$ symmetric and both of its eigenvalues are shifted by an imaginary constant,  
\begin{align}
\label{eigvals_1}
\Lambda_\pm &= \omega_0 -i\frac{\kappa_1+\kappa_2-2\Gamma\delta}{2}\nonumber\\
&\pm \sqrt{\lambda^2(1-\delta) - \left(\frac{\kappa_2-\kappa_1+2\Gamma\delta}{4}\right)^2}
\end{align} 
 However, the exceptional point degeneracy at 
\begin{align}
\label{lamda_EP}
\lambda_{EP}=\frac{|\kappa_2-\kappa_1+2\Gamma\delta|}{4\sqrt{1-\delta}}
\end{align} 
 is still present, and the transition across it is manifest by the two non-orthogonal eigenmodes of $\mathbf{H}_\mathrm{eff}$ having different decay (or amplification) rates at weak coupling $\lambda<\lambda_{EP}$, and a common decay (or amplification) rate at strong coupling, $\lambda>\lambda_{EP}$~\cite{roberto2018,Li2019}. Thus, the key features of the $\mathcal{PT}$ symmetry breaking transition, including the coalescence of the eigenmodes, survives even if the constraint in Eq.(\ref{eq:realchar}) is not fulfilled~\cite{Naghiloo2019}.

Lastly, we discuss the limitations of our analysis. Our effective Hamiltonian $\mathbf{H}_\mathrm{eff}$ is obtained via a linearized theory that works only when the number of photons in the two cavities are not too large. Whenever at least one the eigenvalues of $\mathbf{H}_{eff}$ has a positive imaginary part, the linearized theory predicts an exponential growth for the photon number associated with that eigenmode. This can happen either because $\mathbf{H}_\mathrm{eff}$ is $\mathcal{PT}$ symmetric and $\lambda>\lambda_{EP}$, or because the net gain resulting from the DQD in the first cavity exceeds the loss in the second cavity, i.e. $2\Gamma\delta>\kappa_1+\kappa_2$. Either of these scenarios essentially point to breakdown of the linearized theory. Nevertheless, their identification is important because such behavior in the linearized theory points to onset of lasing in the actual experiment. Though not necessary, a  $\mathcal{PT}$-transition is thus intimately linked with onset or suppression of lasing in the two coupled cavities.  

In the following, we will be looking at the dynamics of expectation values of the field operators or complex quadratures $\langle\hat{b}_\ell(t)\rangle$ ($\ell=1,2$), and those of the cavity-field bilinears like cavity photon numbers $\langle\hat{n}_\ell(t)\rangle=\langle\hat{b}_\ell^\dagger(t)\hat{b}_\ell(t)\rangle$, and the intercavity photon current $I(t)=\lambda\textrm{Im}\langle\hat{b}^\dagger_2(t)\hat{b}_1(t)\rangle$. The dynamics of these quantities can be obtained from Eq.(\ref{eq:eom}). The formal solution of Eq.(\ref{eq:eom}) is given by
\begin{align}
\label{formal_solution}
\left(
\begin{array}{c}
\hat{b}_1 (t)\\
\hat{b}_2 (t)\\
\end{array}
\right) &=  e^{- i\mathbf{H}_{\mathrm{eff}}t}\left(
\begin{array}{c}
\hat{b}_1 (0)\\
\hat{b}_2 (0)\\
\end{array}
\right)\nonumber \\
&-i \int_{0}^t dt^\prime e^{-i\mathbf{H}_{\mathrm{eff}}(t-t^\prime)} \hat{\xi}_A(t^\prime)
\left(
\begin{array}{c}
1\\
0\\
\end{array}
\right) \nonumber \\
&+\frac{E_0e^{-i\omega_d t}\left(1-e^{-i\left(\mathbf{H}_{\mathrm{eff}}-\omega_d\mathbb{I}\right)t  }\right)}{\mathbf{H}_{\mathrm{eff}}-\omega_d\mathbb{I}}
\left(
\begin{array}{c}
0\\
1\\
\end{array}
\right),
\end{align}
where $\mathbb{I}$ is the 2$\times$2 identity matrix. The dynamics of the complex quadratures are obtained by taking expectation value of the above equation. The dynamics of the cavity bilinears are obtained from dynamics of the 
 2$\times2$ equal-time correlation matrix, whose elements are given by  
\begin{align}
\label{def_C}
&\mathbf{C}_{\ell\ell'}(t)=\langle\hat{b}^\dagger_\ell(t)\hat{b}_{\ell'}(t)\rangle.
\end{align}
The expression for $\mathbf{C}(t)$ is obtained by taking the transpose of Eq.(\ref{formal_solution}) and multiplying on the left by its Hermitian conjugate. Assuming that the initial state of the cavities is a coherent state, which satisfies, $\langle\hat{b}^\dagger_\ell(0)\hat{b}_{\ell'}(0)\rangle=\langle\hat{b}^\dagger_\ell(0)\rangle\langle\hat{b}_{\ell'}(0)\rangle$, the expression for the correlation matrix can be written as,
\begin{align}
\label{C_dynamics}
&\mathbf{C}(t) =\mathbf{C}^{(cl)}(t) \nonumber \\
&+\int_{0}^{t} dt^\prime \int_{0}^{t} dt^{\prime\prime} e^{+i\mathbf{H}^*_\mathrm{eff}(t-t')} \mathbf{M}(t^\prime,t^{\prime\prime}) e^{-i\mathbf{H}^T_\mathrm{eff}(t-t^{\prime\prime})},
\end{align}
where, $\mathbf{H}^*_\mathrm{eff}$ denotes complex conjugate of $\mathbf{H}_\mathrm{eff}$, $\mathbf{H}^T_\mathrm{eff}$ denotes the transpose of $\mathbf{H}_\mathrm{eff}$,   $\mathbf{C}^{(cl)}(t)$ is a 2$\times$2 matrix which contains the uncorrelated part, $\mathbf{C}^{(cl)}_{\ell\ell'}(t)=\langle\hat{b}^\dagger_\ell(t)\rangle \langle\hat{b}_{\ell'}(t)\rangle$, and $\mathbf{M}(t',t'')$ is a 2$\times$2 matrix whose only non-zero element is given by the Fourier transform of the noise power spectrum,
\begin{align}
\label{eq:M11}
\mathbf{M}_{11}(t',t'')=\int_{-\infty}^\infty \frac{d\omega}{2\pi} P(\omega) e^{i\omega(t-t^\prime)}.
\end{align}
In absence of quantum fluctuations, we would have $\mathbf{C}(t)=\mathbf{C}^{(cl)}(t)$, which would be consistent with the classical predictions. Thus the connected part of the correlation functions, given by $\mathbf{C}(t)-\mathbf{C}^{(cl)}(t)$ embody the quantum fluctuations. We see from Eq.(\ref{eq:noisevariance}) that $P(\omega)\propto \langle\hat{N}_1 \rangle_{ss}$, which is the steady state occupation of the higher energy mode of the DQD. If the DQD is configured to act as a gain medium, it must be population inverted, which means, $\langle\hat{N}_1 \rangle_{ss}\sim O(1)$. Thus, the quantum fluctuations cannot be neglected and, as we will see below, have non-trivial effects on dynamics of cavity-field bilinears.  


\section{Dynamics of the balanced gain-loss cavities}
\label{Sec: Balanced gain-loss}

 In this section, we look at the the dynamics of cavity observables when the set-up is tuned so that the Hamiltonian $\mathbf{H}_{eff}$ is $\mathcal{PT}$-symmetric, i.e, Eq.(\ref{eq:realchar}) is satisfied. We ensure this by fixing
\begin{align}
\label{parameters2}
\kappa_1=\kappa_2=2~\textrm{MHz},~\varepsilon=7.760~\textrm{GHz},~t_c=0.973~ \textrm{GHz}
\end{align} 
along with the parameters in Table \ref{Table:parameters}.  We assume that initially the second, lossy cavity is empty and the first, DQD-gain cavity is in a coherent state,
\begin{align}
\label{initial_condition}
&\langle\hat{b}_1(0)\rangle = \alpha,\nonumber\\
&\langle\hat{b}_1^\dagger(0)\hat{b}_1(0) \rangle = \alpha^2.
\end{align}
We further assume in this section that there is no additional coherent drive, i.e, $E_0=0$. Satisfying Eq.(\ref{approx2}) under these conditions require $\alpha\gg g\langle \hat{A}_2^\dagger \hat{A}_1 \rangle_{ss}/\omega_0\sim 10^{-4}$ for our choice of parameters.
Under these conditions, we calculate the dynamics complex quadratures, cavity photon numbers $\langle\hat{n}_\ell(t)\rangle=\langle\hat{b}_\ell^\dagger(t)\hat{b}_\ell(t)\rangle$ and the intercavity photon current $I(t)=\lambda\textrm{Im}\langle\hat{b}^\dagger_2(t)\hat{b}_1(t)\rangle$. 

\begin{figure}[!t]
\includegraphics[width=\columnwidth]{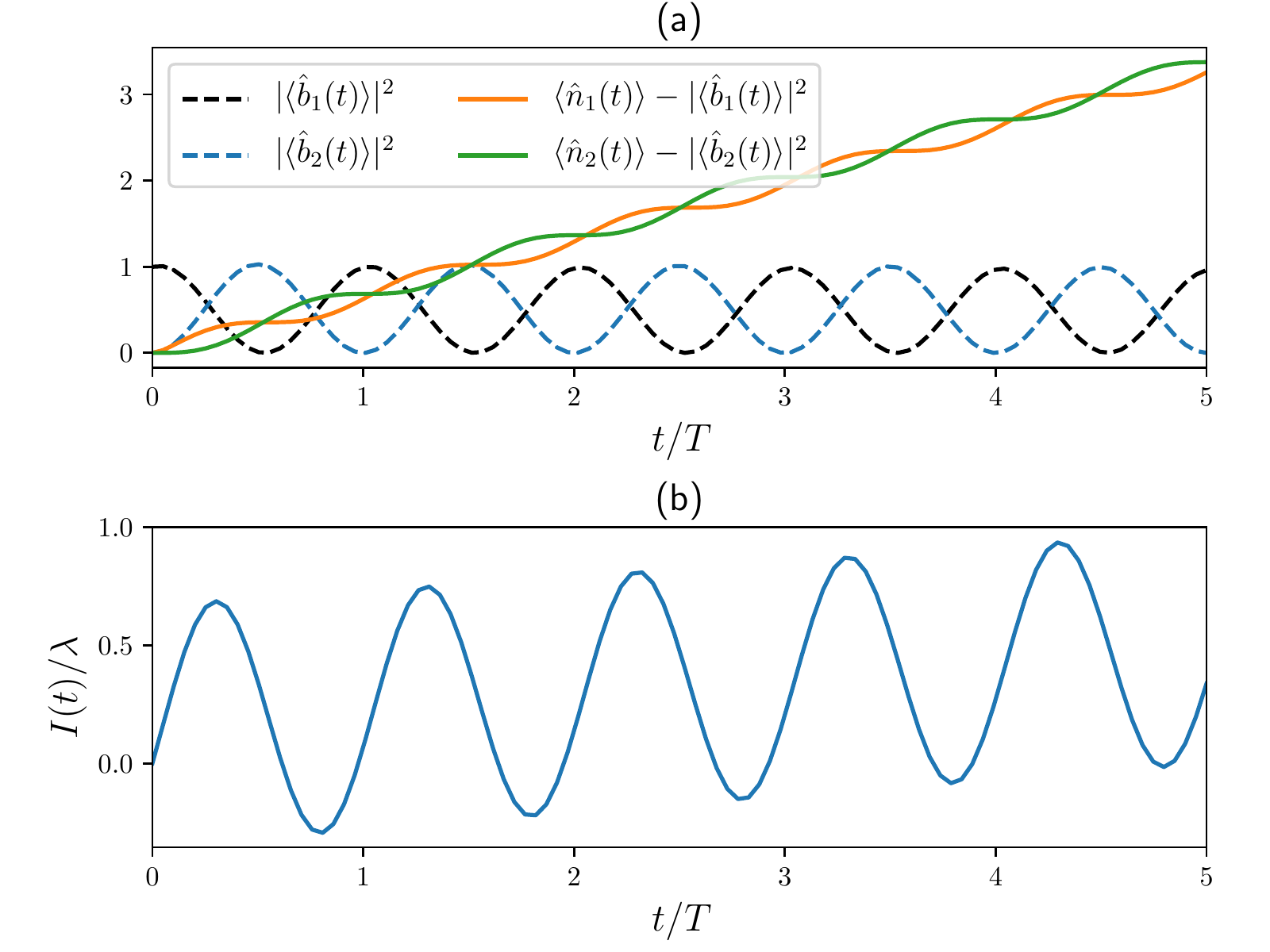} 
\caption{(color online) Oscillatory dynamics of the cavity operator expectation values in the $\mathcal{PT}$-symmetric regime ($\lambda>\lambda_{EP}$). The initial condition is given in Eq.(\ref{initial_condition}). (a) The quadrature magnitudes show oscillations with period $T_\Lambda=2\pi/\left(\Lambda_+-\Lambda_-\right)$. On the other hand, quantum fluctuations, determined by diagonal elements of the matrix $\mathbf{C}(t)$, show a linear growth with superimposed oscillations with the same period. (b) Dynamics of the photonic current $I(t)$, determined by the off-diagonal elements of the matrix $\mathbf{C}(t)$ also shows linear growth with superimposed oscillations. These results are for $\alpha=1$, $\lambda=10$MHz, and other parameters as in Table~\ref{Table:parameters} and Eq.(\ref{parameters2}). The corresponding oscillation period is $T_\Lambda=320$ ns. }
\label{fig:PT_dynamics} 
\end{figure}

The most remarkable effect of having $\mathcal{PT}$ symmetry, Eq.(\ref{eq:pt}), for the Hamiltonian $\mathbf{H}_\mathrm{eff}$ is the possibility that an open system shows periodic dynamics when $\lambda>\lambda_{EP}$, i.e, when the cavities are strongly coupled. Figure~\ref{fig:PT_dynamics}(a) shows the numerically obtained results for the amplitude of the complex quadratures with dashed lines. We get oscillatory dynamics with period $T_\Lambda=2\pi/\left(\Lambda_+-\Lambda_-\right)$. Thus, they indeed show the signatures of a $\mathcal{PT}$-symmetric phase. In contrast, when it comes to cavity-field bilinears, i.e. photon numbers and the intercavity current, the oscillatory behavior is superseded by the effects of quantum fluctuations at long times.  

To see this surprising result, we note that, away from the exceptional point, the non-Hermitian Hamiltonian can be diagonalized by a similarity transformation. Let $\mathbf{H}^T_\mathrm{eff}=\mathbf{R}\mathbf{\Lambda}\mathbf{R}^{-1}$, where $\mathbf{R}$ has right eigenvectors of $\mathbf{H}^T_\mathrm{eff}$ and $\mathbf{\Lambda}$ is a 2$\times$2 diagonal matrix with real eigenvalues $\Lambda_\pm$. In this skewed basis, the elements of the correlation matrix $\tilde{\mathbf{C}}(t)=\mathbf{R}^{-1}\mathbf{C}(t)\mathbf{R}$ can be explicitly calculated. For times $t\gg\Gamma^{-1}$, we get 
\begin{align}
\label{C_tilde_soln1}
\tilde{\mathbf{C}}_{\ell \ell} (t) &=\tilde{\mathbf{C}}_{\ell \ell} (0)+\frac{2g^2\langle\hat{N}_1\rangle_{ss}\tilde{m}_{\ell\ell}}{\Gamma^2}\left\{\frac{\Gamma t}{1+(\left(\Lambda_+-\Lambda_-\right)/2\Gamma)^2}\right.\nonumber\\
&-\left.\frac{1-(\left(\Lambda_+-\Lambda_-\right)/2\Gamma)^2}{[1+(\left(\Lambda_+-\Lambda_-\right)/2\Gamma)^2]^2}\right\},\\
\tilde{\mathbf{C}}_{12}(t) &= e^{i\left(\Lambda_+-\Lambda_-\right) t}\tilde{\mathbf{C}}_{12}(0)-\frac{g^2\langle \hat{N}_1\rangle_{ss}\tilde{m}_{12}}{\Gamma^2}\times\nonumber \\
\label{C_tilde_soln2}
&\Big\{\frac{-2i\Gamma}{\left(\Lambda_+-\Lambda_-\right)}\frac{(1-e^{i\left(\Lambda_+-\Lambda_-\right) t})}{1+(\left(\Lambda_+-\Lambda_-\right)/2\Gamma)^2}\nonumber \\
&+ \frac{1+e^{i\left(\Lambda_+-\Lambda_-\right) t}}{1+(\left(\Lambda_+-\Lambda_-\right)/2\Gamma)^2}\Big\}.
\end{align}
where $\tilde{m}_{\ell\ell'}=\left(\mathbf{R}^{-1}(\mathbbm{1}_2+\sigma_z)\mathbf{R}/2\right)_{\ell\ell'}$. We see from Eqs.(\ref{C_tilde_soln1})-(\ref{C_tilde_soln2}) that while $\tilde{C}_{12}(t)$ oscillates with period $T_\Lambda$, the diagonal elements $\tilde{C}_{\ell \ell}(t)$ grow linearly with time due to the quantum fluctuations arising from the DQD. This linear growth in the eigenmode occupation numbers will eventually make the assumption $g \sqrt{n_\mathrm{photons}} \leq O(\Gamma)$ (weak cavity-DQD coupling) invalid. So, starting in the $\mathcal{PT}$-symmetric region, due to quantum fluctuations the system will be eventually driven out of the region of validity of a $\mathcal{PT}$-symmetric effective Hamiltonian description. For our choice of parameters in Table~\ref{Table:parameters}, $g \sqrt{n_\mathrm{photons}} \leq O(\Gamma)$ corresponds to $n_\mathrm{photons}\lesssim 50$.

To explicitly show the effect of quantum fluctuations, we plot  $\langle\hat{n}_\ell(t)\rangle-|\langle\hat{b}_\ell(t)\rangle|^2$ (solid lines) in Fig.~\ref{fig:PT_dynamics}(a). While the amplitudes of the complex quadratures show perfect periodic oscillations of period $T_\Lambda$, the quantum fluctuations show small oscillations about a linear growth, as discussed above. Fig.~\ref{fig:PT_dynamics}(b) shows dynamics of the photonic current $I(t)/\lambda$. This also shows a linear growth but with a much smaller slope, along with large oscillations of period $T$. These larger oscillations and weaker linear growth in the photon current is due to the larger weight of the oscillatory off-diagonal elements $\tilde{\mathbf{C}}_{12}(t)$, Eq.(\ref{C_tilde_soln2}), and smaller weight of the linearly growing diagonal elements $\tilde{\mathbf{C}}_{\ell\ell}(t)$, Eq.(\ref{C_tilde_soln1}). 

\begin{figure}[!t]
\includegraphics[width=\columnwidth]{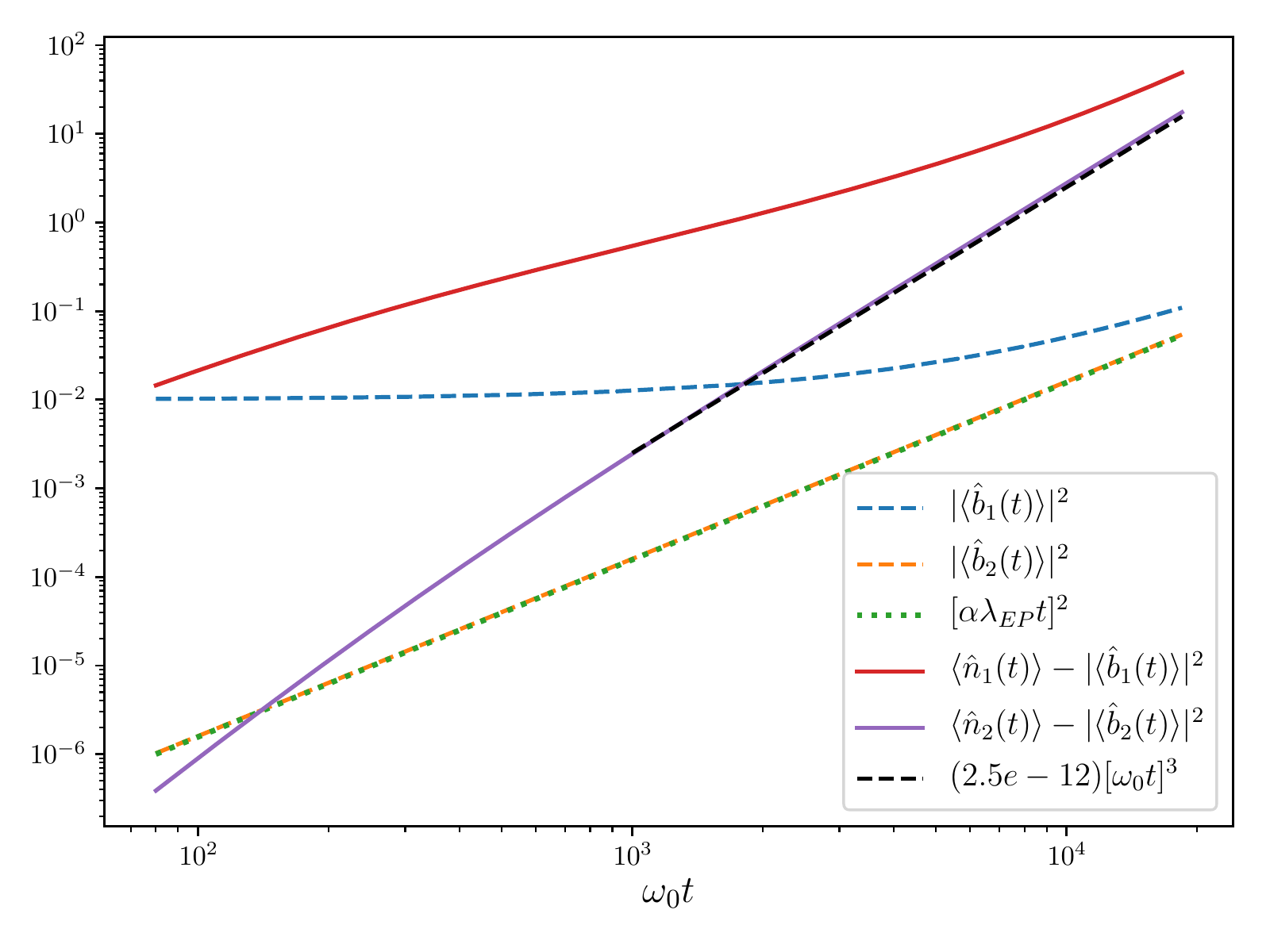} 
\caption{(color online) 
 Power-law dynamics of the cavity operator expectation values at $\lambda=\lambda_{EP}=\kappa_2/\sqrt{1-\delta}$. The results are for $\alpha=0.1$ and other parameters as in Table~\ref{Table:parameters} and Eq.(\ref{parameters2}), which leads to $\lambda_{EP}=1$ MHz.  The maximum time shown here corresponds to when there are $\sim 50$ photons in the cavities, above which the linearized model cannot be applied for our choice of parameters.  For the initially empty, lossy cavity $|\langle b_2(t)\rangle|^2$ grows quadratically with time, while the quantum fluctuations in quadrature amplitude grow as $t^3$. For the cavity initially in coherent state, $|\langle b_1(t)\rangle|^2$ remains constant at times $t\ll 1/\kappa_2$, and tends to a quadratic growth as $t\gg1/\kappa_2$. The quantum fluctuations in quadrature amplitude of this cavity also tend to a $t^3$ behavior at these times. 
}
\label{fig:PT_dynamics_EP} 
\end{figure}

Next, let us see the dynamics of system at the exceptional point $\lambda=\lambda_{EP}$. At the exceptional point, $\mathbf{H}^T_\mathrm{eff}$ is not digaonalizable, but can be be brought into a Jordan normal form via a similarity transform $S^{-1}\mathbf{H}^T_\mathrm{eff}S~=~\omega_0 (\mathbbm{1}_2 + \sigma_+)$, where $\sigma_+=(\sigma_x+\sigma_y)/2$. Multiplying on the left by $S$ and on the right by $S^{-1}$, we have, $\mathbf{H}^T_\mathrm{eff}=\omega_0(\mathbbm{1}_2+S\sigma_+S^{-1})$.  This means that the Hamiltonian satisfies the characteristic equation $(\mathbf{H}^T_\mathrm{eff}~-~\omega_0\mathbbm{1}_2)^2~=~0$, at this point. So, the time-evolution operator Taylor expansion terminates at first order, i.e. 
\begin{align}
&e^{-i\mathbf{H}^T_\mathrm{eff}t}=e^{-i\omega_0t}\left[(1+i\omega_0t)\mathbbm{1}_2-i\mathbf{H}^T_\mathrm{eff}t\right]. 
\end{align}
Therefore, with our initial conditions, the time-dependent complex quadratures become 
\begin{align}
\label{eq:b1quad}
& \langle \hat{b}_1(t) \rangle = \alpha e^{-i\omega_0t} \left(1+\kappa_2 t/2\right),\\
\label{eq:b2quad}
& \langle \hat{b}_2(t) \rangle = -i \alpha e^{-i\omega_0t}(\lambda_{EP}t).
\end{align}
Thus, the empty, lossy cavity has a quadratic growth, i.e. $|\langle \hat{b}_2(t) \rangle|^2\propto t^2$. In the gain cavity,  $|\langle \hat{b}_1(t) \rangle|^2$ remains flat for small times $t\ll 1/\kappa_2$, but switches to quadratic growth at times $t\gg1/\kappa_2$. Similarly, it can be shown that at long times $t\gg1/\kappa_2$, photon numbers in both cavities, $\langle\hat{n}_1(t)\rangle$ and $\langle\hat{n}_2(t)\rangle$ scale as $t^3$. The numerically obtained results for the quadrature magnitude (squared) (dashed lines) and quantum fluctuations (solid lines) are shown in Fig.~\ref{fig:PT_dynamics_EP}. The dotted line shows the quadratic fit to the empty-cavity $|\langle b_2(t)\rangle|^2$, Eq.(\ref{eq:b2quad}), while the dot-dashed line shows the cubic fit for the quantum fluctuation results. We remind the reader that our model remains valid only at times when $n_\mathrm{photons}\lesssim 50$, and that determined the time-range chosen in Fig.~\ref{fig:PT_dynamics_EP}. 

Lastly, we consider the system's behavior in the $\mathcal{PT}$-symmetry broken region, i.e for  $\lambda<\lambda_{EP}$. Here, due to the presence of the amplifying eigenmode that is delocalized over the two sites, the photon numbers in both gain and loss cavities grow exponentially at small times, and provide a short-time cutoff beyond which the linearized theory fails. As mentioned before, identifying this regime is important because, such behavior in the linearized system points to onset of lasing in the actual experimental set-up. 

\section{Signatures of exceptional points for dissipative cavities}
\label{Sec: Dissipative PT}

In the previous section, we have seen the dynamics of the cavities, including the effects of quantum fluctuations,  when the gain from the DQD-unit is balanced by the losses of the cavities, Eq.(\ref{eq:realchar}). In this section, we look at dissipative cavities, where the combined loss from the two cavities exceeds the DQD gain in the first cavity, i.e 
\begin{align}
\kappa_1+\kappa_2>2\Gamma\delta.
\end{align}
 Even under this condition, this system can traverse across exceptional points on tuning either the coupling between the cavities, or one of the losses of one of the cavities.  We show that this can be observed in particular input-output experiments through transmission, phase response and fluctuations, and may lead to interesting applications like loss-induced lasing. Note that, in-situ tuning of coupling between cQED resonators, as well as, of the resonator losses has already been demonstrated experimentally \cite{EP_superconducting_circuits_expt_2019,Tunable_loss_expt_2013,
Tunable_loss_expt_2014,Tunable_loss_expt_2016,
Tunable_coupling_expt_2015}.

\subsection{Individually lossy cavities}
\label{subsec:ilc}

First, we consider the case where each cavity is individually lossy, with same dissipation rate. In particular, we will assume, starting from a $\mathcal{PT}$-symmetric condition, the dissipation rate in each cavity is increased by the same amount $\kappa$. Therefore, the effective non-Hermitian Hamiltonian in Eq.(\ref{eq:Heff}) changes to
\begin{align}
&\mathbf{H}_{\mathrm{eff}}=\left(\omega_0-i\frac{\kappa}{2}\right)\mathbbm{1}_2+\frac{\lambda}{2}(2-\delta)\sigma_x-i\frac{\lambda\delta}{2}\sigma_y+i\frac{\kappa_2}{2}\sigma_z, 
\end{align}
while Eq.(\ref{eq:realchar}) is still satisfied for $\kappa_1$ and $\kappa_2$. The eigenvalues of the effective Hamiltonian now have an additive imaginary part $\Lambda_{\pm}\rightarrow\Lambda_{\pm}-i{\kappa}/{2}$, where $\Lambda_{\pm}$ are given in Eq.(\ref{Lamda_pm_balanced_gain_loss}). However, the real parts of the eigenvalues remain the same. The overall dissipation will cause the system to reach a time-independent non-equilibrium steady state. The system will show a passive $\mathcal{PT}$-transition as a function of $\lambda$, with $\lambda_{EP}$ still given by Eq.(\ref{lamda_EP_balanced_gain_loss}), where the real parts bifurcate \cite{roberto2018,Li2019,Naghiloo2019}.  In what follows, we show that, when $\kappa=2\kappa_2$, properties of this steady state capture the passive $\mathcal{PT}$-transition. This is accomplished via a standard input-output experiment where a weak, coherent drive $E_0 e^{-i\omega_d t}$ with amplitude $E$ and frequency $\omega_d$ is applied to one of the cavities and the steady-state transmitted signals are observed in either cavity. We will assume that the coherent drive is in the cavity without the DQD in  it. We first go to the rotating frame with respect to the drive frequency, $\hat{b}_\ell^{rot}(t)e^{i\omega_d t}=\hat{b}_\ell(t)$. The transmitted signals $T_\ell$ at the two cavities are given by the expectation value of the long-time solution of Eq.(\ref{formal_solution}) scaled properly by the  dissipation rates and $E_0$, 
\begin{align}
&\left[
\begin{array}{c}
T_1/(\kappa_1+\kappa) \\
T_2/(\kappa_2+\kappa) \\
\end{array}
\right]
=\lim_{t\rightarrow\infty}\frac{1}{E_0}\left[
\begin{array}{c}
\langle \hat{b}^{rot}_1 \rangle\\
\langle \hat{b}^{rot}_2 \rangle\\
\end{array}
\right]\nonumber \\
&=i(\mathbf{H}_\mathrm{eff}-\omega_d\mathbbm{1}_2)^{-1}\left[
\begin{array}{c}
0 \\
1\\
\end{array}
\right].
\end{align}
\begin{figure}[t]
\includegraphics[width=\columnwidth]{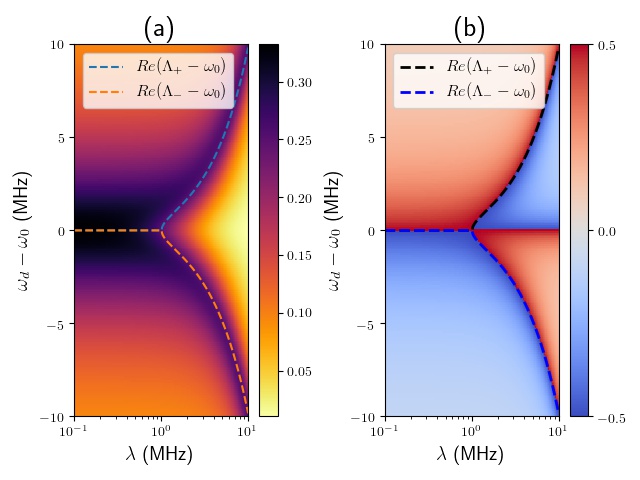} 
\caption{(color online) (a) The transmission amplitude in the lossy cavity with coherent drive $|T_2|$ as a function of drive frequency $\omega_d$ and the coupling between the cavities $\lambda$ shows clear signature of the exceptional point $\lambda_{EP}$ through bifurcation of the transmission peak. (b) The phase response of the lossy cavity $\phi_2$ in units of $\pi$ is plotted as function of $\omega_d$ and $\lambda$. The real parts of eigenvalues of $(\mathbf{H}_\mathrm{eff}-\omega_0\mathbbm{1}_2)$ are plotted as dashed lines in both plots. It is clear from both panels that the passive $\mathcal{PT}$-transition is detectable in an input-output experiment. Parameters $\kappa_1=\kappa_2=2$ MHz, $\kappa=4$ MHz  $\varepsilon=7.760$ GHz and $t_c=0.973$ GHz and others the same as in Table.~\ref{Table:parameters}. }
\label{fig:input_output} 
\end{figure}
We write the transmitted signal at $\ell$th cavity as $T_\ell=|T_\ell|e^{i\phi_{\ell}}$, where $|T_\ell|$ is the amplitude of the transmitted signal, and $\phi_{\ell}$ is the phase response, $\phi_\ell \in [-\pi/2,\pi/2]$. Both of the transmission amplitude and the phase response can be experimentally measured \cite{Kontos_DQD_cQED_review_2,wallraff2019_1,wallraff2019_2,
probing_electron_phonon_DQD_cQED_1}. Explicit expressions for both the amplitude and the phase response can be obtained (see Appendix.~\ref{transmission_results}).  When $\lambda>\lambda_{EP}$ and $\omega_d=\omega_0 \mp \left(\Lambda_+-\Lambda_-\right)$, we observe that the phase response of the second cavity is given by
\begin{align}
\label{phi_2_individually_lossy}
\phi_2 = \tan^{-1}\left(\frac{\kappa_2\pm 4\left(\Lambda_+-\Lambda_-\right)}{\pm\left(\Lambda_+-\Lambda_-\right) \left(2\kappa_2-\kappa\right)}\right)
\end{align}
Thus, when $\kappa=2\kappa_2$, the phase response becomes $\pm \pi/2$. This means, there will be a phase change of $\pm\pi$ when $\omega_d$ is tuned across values equal to $\omega_0 \mp \left(\Lambda_+-\Lambda_-\right)$. Further, it can also be checked that $\textrm{det}(\mathbf{H}_\mathrm{eff}-\omega_d\mathbbm{1}_2)$ has a minimum at $\omega_d=\omega_0 \mp \left(\Lambda_+-\Lambda_-\right)$ under this condition.  So, when $\kappa=2\kappa_2$, both the transmission amplitude and the phase response of the lossy cavity will accurately capture the bifurcation of real parts of eigenvalues of $\mathbf{H}_\mathrm{eff}$ across a passive $\mathcal{PT}$ transition. This is shown in Fig.~\ref{fig:input_output}, where both the transmission amplitude $|T_2|$ and the phase repsonse $\phi_\ell$, detected at the lossy, second cavity as a function of $\omega_d$ and $\lambda$ are shown with color coding. The real parts of the eigenvalues are also plotted as function of $\lambda$ for comparison. The passive $\mathcal{PT}$ transition and the location of the exceptional point  are completely clear both in the transmission amplitude and in the phase response. The transmission amplitude peaks when the drive frequency $\omega_d$ is equal to the real parts of the eigenvalues of $\mathbf{H}_\mathrm{eff}$, while the phase response undergoes a change of $\pi$ when  $\omega_d$ is changed across these values.  


Note that even without the gain medium (i.e, in absence of the DQD), the passive $\mathcal{PT}$-transition in lossy cavities can be seen, through input-output experiments, if the two losses are different. In Eq.(\ref{def_H_eff}), this corresponds to setting $\delta=0$. However, the difference of that situation with the one with a gain will be seen in the quantum fluctuations of the complex quadratures. As we will see below, the steady-state photon number will encode quantum fluctuations from the DQD-gain, thus distinguishing between a purely lossy set-up and the set-up with a gain DQD.
\begin{figure}[!t]
\includegraphics[width=\columnwidth]{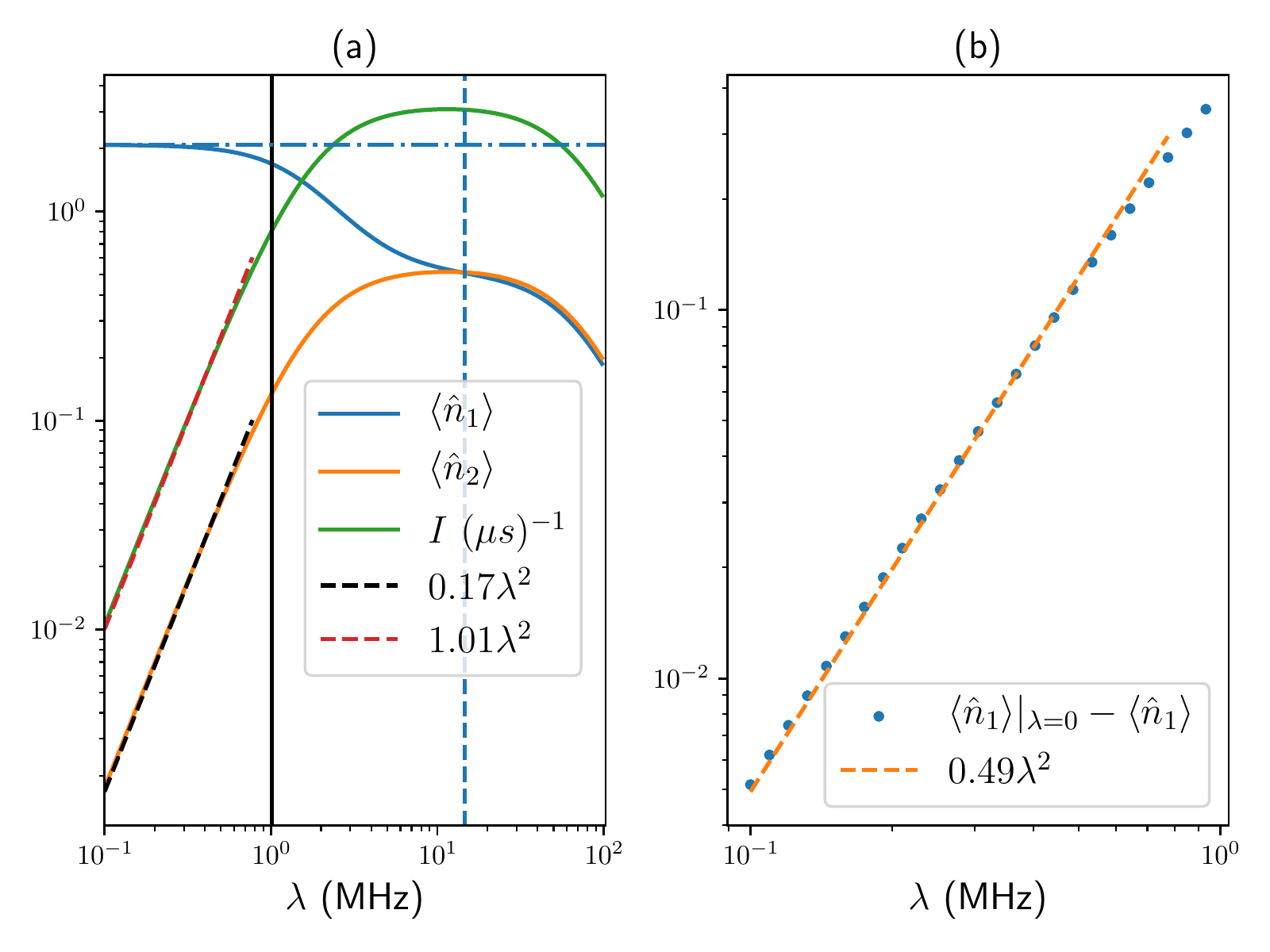} 
\caption{(color online) (a) Steady-state values of photon occupation numbers of the left ($\langle\hat{n}_1\rangle$) and the right ($\langle\hat{n}_2\rangle$) cavities, as well as the photon current ($I$) from the left to the right cavity as a function of $\lambda$ for the dissipative cavities in absence of any additional coherent drive. The horizontal dash-dotted line is $\langle\hat{n}_1\rangle_{\lambda=0}$. The position of the exceptional point, $\lambda_{EP}=1$ MHz is shown by the vertical line, and the vertical dashed line is at $\lambda=g$, where $g$ is the effective coupling between the left cavity and the DQD. (b) Plot of the $\langle\hat{n}_1\rangle|_{\lambda=0}-\langle\hat{n}_1\rangle$ with $\lambda$ for $\lambda<\lambda_{EP}$. System parameters are the same as in Fig.~\ref{fig:input_output}.}
\label{fig:dissipative_steady_state} 
\end{figure}

To see the effect of quantum fluctuations, as before, we will be interested in the steady-state values, $\langle\hat{n}_\ell\rangle-|\langle\hat{b}_\ell\rangle|^2$ ($\ell=1,2$).  As can be seen from Eq.(\ref{C_dynamics}), the quantum fluctuations are independent of the presence of the coherent drive.  So it suffices to look at the case where the coherent drive is absent. In that case, because the overall cavities are dissipative, expectation values of the field operators in steady state are zero. So, we look at the steady-state photon numbers $\langle\hat{n}_\ell\rangle$, and the photon current $I$ as a function of $\lambda$. We numerically calculate these quantities using Eq.(\ref{C_dynamics}), and obtain the result at time $t\gg1/\kappa$. These results are shown in Fig.~\ref{fig:dissipative_steady_state}. 

At zero coupling, there are no steady-state photons in the second cavity and there is no photon current. However, due to the presence of the DQD in the left cavity there is a steady-state photon occupation in the first cavity when $\kappa_1+\kappa>2\Gamma\delta$. As the coupling is increased to values $\lambda<\lambda_{EP}$, steady-state values photon number in the second cavity and the photon current both increase quadratically with $\lambda$. The quadratic growth slows down and the photon number in the DQD-cavity suppressed when $\lambda\rightarrow\lambda_{EP}$,  For $g\gtrsim\lambda\gtrsim\lambda_{EP}$, where $g$ is the effective coupling between the DQD and the first cavity, the photon numbers in both the cavities approach the same value, and photon current approaches maximum (Fig.~\ref{fig:dissipative_steady_state}a). As $\lambda$ is further increased, the occupation of both cavities still remains almost the same, but they decrease with $\lambda$. The current also decreases correspondingly. This is because, in this regime, the relative strength of coupling to the DQD decreases. The suppression of the photon number in the DQD-cavity with increasing $\lambda$ is shown in Fig.~\ref{fig:dissipative_steady_state} b. It, too, shows a quadratic scaling that is expected from a perturbation theory in $\lambda$. We remind the reader that the nozero steady-state values are due to the presence of quantum fluctuations from the DQD. Without them, the dissipative cavities would be empty in the steady state.


\subsection{Loss induced lasing and amplification}
\label{subsec:lil}

In previous subsection, we investigated the passive $\mathcal{PT}$-transition as a function of intercavity coupling $\lambda$. A similar transition can occur if the loss in one of the cavities is increased, giving way to the counterintuitive phenomenon of loss induced lasing~\cite{loss_induced_lasing_expt,Brandstetter2014}. In this subsection, we show that this phenomenon can be observed in our set-up. We further argue that it is a consequence of having a $\mathcal{PT}$-symmetric phase in this set-up.

\begin{figure}[!t]
\includegraphics[width=\columnwidth]{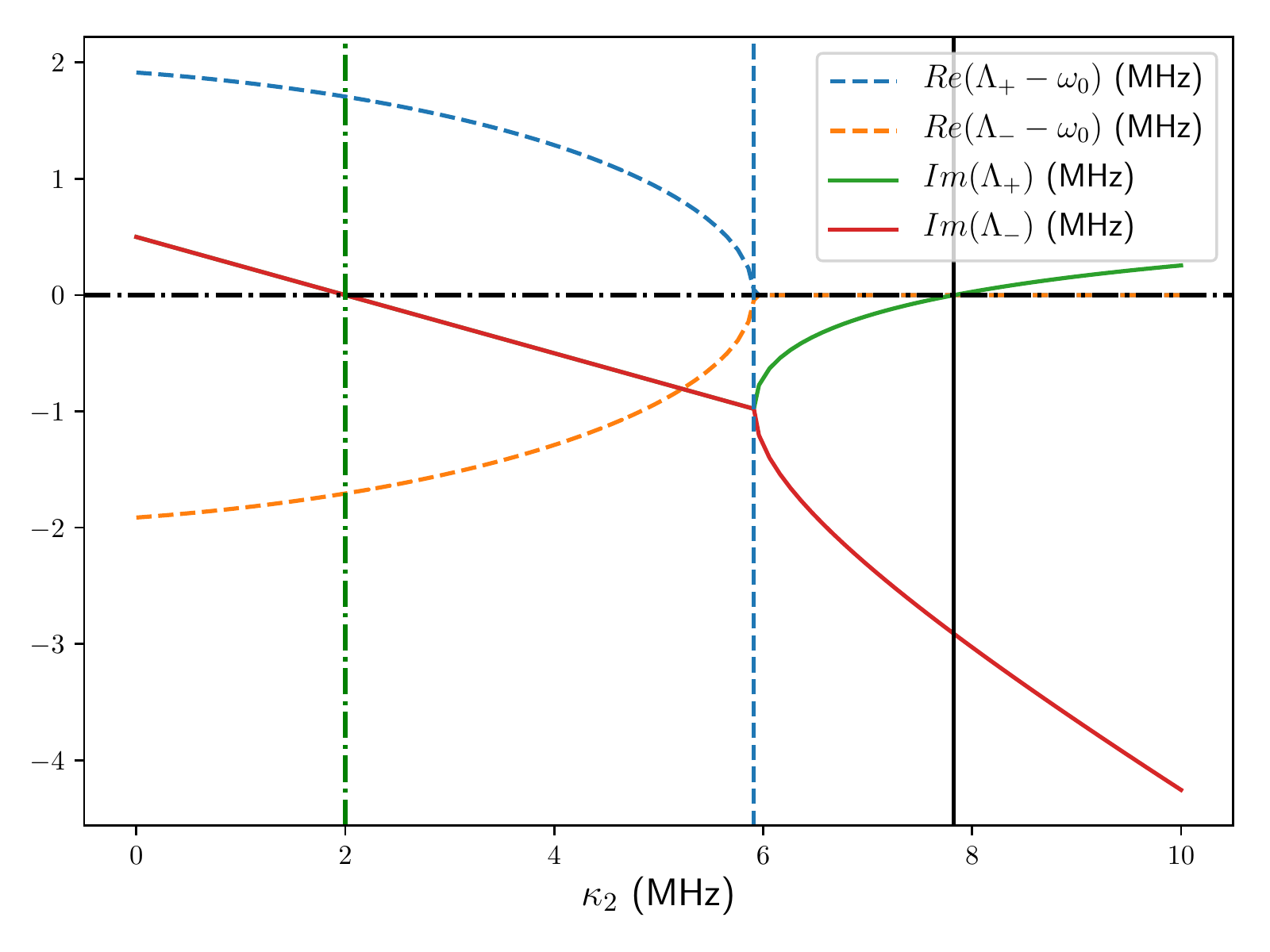} 
\caption{(color online) The real and imaginary parts of the eigenvalues of $\mathbf{H}_{\rm eff}$ are plotted as a function of the loss at the right cavity $\kappa_2$. The vertical continuous black line shows the position of $\kappa_2=\kappa_2^{th}$, where imaginary part of one of the eigenvalues become zero. The vertical dashed blue line shows the exceptional point $\kappa_2=\kappa_2^{EP}$. The vertical dash-dotted green line shows $\kappa_2=2\Gamma \delta-\kappa_1$, which is balanced-gain-loss condition. For $\kappa_2>\kappa_2^{th}$, the system gets an overall gain pointing to onset of loss-induced lasing.  Parameters: $\lambda=2$MHz, $\kappa_1=2$MHz, $\varepsilon=7.760$ GHz and $t_c=0.973$ GHz. Other parameters are same as in Table.~\ref{Table:parameters}.}
\label{fig:PT_transition_kappa} 
\end{figure}

\begin{figure}[!t]
\includegraphics[width=\columnwidth]{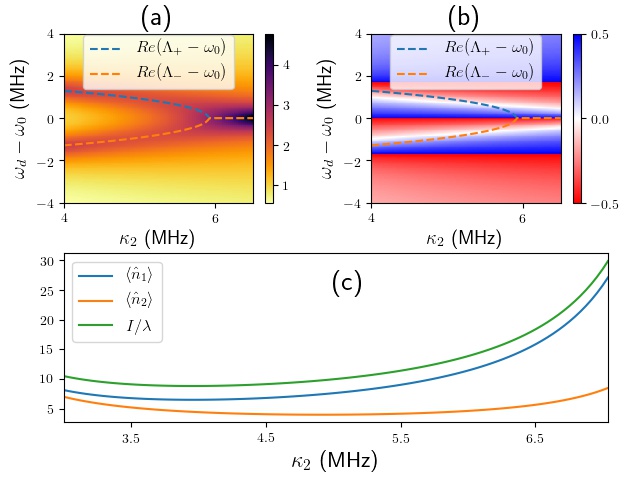} 
\caption{(color online) The top panel figure shows in color code (a) amplitude $\mid T_2\mid$ and (b) phase response $\phi_2$ of the transmitted signal at the lossy cavity for an input-output experiment, as a function of the loss at the right cavity $\kappa_2$, and drive frequency $\omega_d$. The phase response is given in units of $\pi$. Here, the input drive is at the right cavity. The $\mathcal{PT}$-transition on increasing $\kappa_2$ is approximately tracked by the amplitude response of the input-output experiment, but the phase response does not tack the transition.   The real parts of the eigenvalues are also shown by dashed lines for comparison. The bottom panel (c) shows the average steady state photon numbers of the two cavities and the photon current as a function of $\kappa_2$, in absence of any coherent drive.  Parameters are same as in Fig.~\ref{fig:PT_transition_kappa}.}
\label{fig:input_output_kappa} 
\end{figure}

To see this, we consider the loss $\kappa_2$ in the second cavity as the tunable parameter, whereas the rest of the set-up is kept fixed.  Note that, transition across exceptional point via tuning the loss in coupled cQED resonators has been recently shown experimentally \cite{EP_superconducting_circuits_expt_2019}, but in absence of any gain medium. For concreteness, in our set-up, let us consider that $\kappa_2$ is adiabatically increased starting from a balanced gain-loss $\mathcal{PT}$-symmetric condition. On slight increase of $\kappa_2$ the system becomes dissipative and approaches a steady state in the long time limit. However, as $\kappa_2$ is increased further, from Eq.(\ref{lamda_EP}), we see that, an exceptional point is reached when 
\begin{align}
\label{kappa_EP}
&\kappa_2=\kappa_2^{EP}=\kappa_1-2\Gamma\delta+2\lambda\sqrt{1-\delta}.
\end{align} 
For $\kappa_2<\kappa_2^{EP}$, $\left(\Lambda_+-\Lambda_-\right)$ is real (see Eq.(\ref{eigvals_1})), and hence, the imaginary parts of the eigenvalues of $\mathbf{H}_{\rm eff}$ are same. But for $\kappa_2>\kappa_2^{EP}$, $\left(\Lambda_+-\Lambda_-\right)$ is imaginary. In this case, the imaginary parts of the two eigenvalues of $\mathbf{H}_{\rm eff}$ bifurcate into two different values. On further increasing $\kappa_2$, the system remains dissipative as long as
\begin{align}
\label{kappa_th}
&\kappa_2+\kappa_1-2\Gamma\delta > \sqrt{\left(\kappa_2-\kappa_1+2\Gamma\delta\right)^2-4\lambda^2(1-\delta)}\nonumber \\
&\Rightarrow \kappa_2 < \kappa_{2}^{th},~~ \kappa_{2}^{th}=\frac{4\lambda^2 (1-\delta)}{2\Gamma\delta-\kappa_1}.
\end{align}
If $\kappa_2$ is increased beyond $\kappa_2^{th}$, imaginary part of one of the eigenvalues of $\mathbf{H}_{eff}$ become positive. Thus, starting from a  condition when the overall system is dissipative,  on increasing the loss of the right cavity beyond the threshold value, the system gets an overall effective gain, which points to onset of lasing. This is the rather counter-intuitive phenomenon of loss induced lasing that occurs in this set-up. It can be shown that this is a consequence of existence of a $\mathcal{PT}$-symmetric phase with balanced-gain loss.  To see this, we note that the exceptional point $\kappa_2^{EP}$ given in Eq.(\ref{kappa_EP}) is only possible if $\lambda~>~(2\Gamma\delta-\kappa_1)/(2\sqrt{1-\delta})$. Under balanced gain loss, $\kappa_2=2\Gamma\delta-\kappa_1$, and this regime of $\lambda$ corresponds to the $\mathcal{PT}$-symmetric phase (see Eq.(\ref{lamda_EP_balanced_gain_loss})). This shows that, it was necessary that we started adiabatically increasing $\kappa_2$ from the $\mathcal{PT}$-symmetric phase of the balanced gain-loss system.  Further, going back to Eq.(\ref{eigvals_1}), it is possible to see that there is another exceptional point possible with $\kappa_2$ as the only tuning parameter. This occurs when $\kappa_2=\kappa_1-2\Gamma\delta-2\lambda\sqrt{1-\delta}$. Since $\kappa_2,\lambda>0$, occurrence of this exceptional point requires $\kappa_1-2\Gamma\delta>0$. Under this condition, both cavities are lossy, and hence, there can neither be lasing, nor a balanced gain-loss $\mathcal{PT}$-symmetric phase.  This can also be seen from Eq.(\ref{kappa_th}), which shows that a necessary condition to see loss induced lasing in our set-up is  $\kappa_1-2\Gamma\delta<0$. Thus, the phenomenon of loss-induced lasing that can be observed in this set-up is a consequence of having a $\mathcal{PT}$-symmetric phase with balanced gain-loss. (Note that, in other set-ups loss induced lasing can occur without any exceptional point \cite{Longhi_loss_induced_lasing}.) This explicitly requires a gain medium and cannot be observed in the existing experiments in the quantum regime \cite{Xiao2017,Klauck2019, Li2019,Non_hermitian_AB_ring_expt_2020,Wu2019,
Naghiloo2019,EP_superconducting_circuits_expt_2019,
Anti_PT_quantum_experiment_2020,
Non_hermitian_conductance_expt_2019,Passive_PT_critical_phenomena_expt_2019}, none of which features a gain medium.  The real and imaginary parts of the eigenvalues of $\mathbf{H}_{eff}$ as a function of $\kappa_2$ are plotted in Fig.~\ref{fig:PT_transition_kappa}. The positions of $\kappa_2^{EP}$, $\kappa_2^{th}$ and $\kappa_2$ corresponding to the balanced gain-loss $\mathcal{PT}$-symmetric condition are shown.

Our linearized theory allows us to explore the case $\kappa_2<\kappa_2^{th}$. In this case, since the overall system is dissipative, there is a unique time-independent non-equilibrium steady state. As before, we look at an input-output experiment with a coherent drive at the second, lossy cavity.  We look at the amplitude of the transmitted signal as well as the phase response at the lossy cavity. The results for the transmitted signal at the lossy cavity are shown in  Fig.~\ref{fig:input_output_kappa}. Fig.~\ref{fig:input_output_kappa}(a) shows the transmission amplitude $|T_2|$. The first point to note is that $|T_2|>1$. Thus, there is amplification of the transmitted signal in the lossy cavity. The peak in the transmission amplitude as a function of $\omega_d$ and $\kappa_2$ approximately tracks the transition across the exceptional point. Moreover, the amplification of signal at the lossy cavity at resonant drive $\omega_d=\omega_0$ increases with increase in the loss of the cavity. Thus, in this system, we have a novel microwave amplifier \cite{photon_emission_DQD_cQED, Giant_photon_gain_DQD_cQED, DQD_cQED_laser_theory,wallraff2015_1} with loss-induced enhancement of amplification facilitated by transition across an exceptional point.

Fig.~\ref{fig:input_output_kappa}(b) shows the phase response at the lossy cavity. The phase response does not capture the transition across the expectional point. Nevertheless, the phase response does show some interesting features. There occurs a phase change of $\pi$ at specific values of $\omega_d$, independent of $\kappa_2$. These values are given by
\begin{align}
\label{pi_by_2_condition_loss_induced}
\omega_d=\omega_0,~ \omega_0\pm \sqrt{\lambda^2(1-\delta)-(\kappa_1-2\Gamma \delta)^2/4}.
\end{align}
On the other hand, the phase response $\phi_2$ is zero at values which depend on $\kappa_2$. These values are given by
\begin{align}
\label{zero_condition_loss_induced}
&\omega_d=\omega_0 + \frac{\kappa}{2\Gamma\delta-\kappa_1}\nonumber \\
&\pm \sqrt{\left(\frac{\kappa}{2\Gamma\delta-\kappa_1}\right)^2+\left(\frac{\kappa}{2}\right)^2 + \left(\Lambda_+-\Lambda_-\right)^2}, 
\end{align}
where $\kappa=\left(\kappa_2+\kappa_1-2\Gamma\delta\right)/2$. We will like to mention that even though loss induced suppression and revival of lasing has been experimentally observed in classical set-ups previously \cite{loss_induced_lasing_expt,Brandstetter2014}, to our knowledge, the phase response has not been previously measured. This measurement is experimentally possible in our DQD-cQED set-up (see, for example, \cite{Kontos_DQD_cQED_review_2,wallraff2019_1,wallraff2019_2,
probing_electron_phonon_DQD_cQED_1}).  

To see the effect of quantum fluctuations, as before, we plot the steady state photon numbers of the two cavities, and the photon current in absence of any coherent drive in Fig.~\ref{fig:input_output_kappa}(c).  They show non-monotonic behavior with $\kappa_2$. Thus, the photon number of the two cavities and the photon current between the two cavities first decrease but then increase with increase in $\kappa_2$, even in absence of any coherent drive. We remind the reader once again that, since there is overall dissipation, without the effect of quantum fluctuations from the DQD, the steady state photon numbers would be zero in absence of any coherent drive. Thus, the loss-induced increase in average photon number of the cavities in absence of any coherent drive is a feature possible only in a non-Hermitian system with a gain medium having quantum fluctuations. This feature of the non-equilibrium steady state thereby distinguishes our proposed set-up from previous experiments involving gain in classical systems \cite{Ozdemir_2019_review,El-Ganainy_2018_review,Longhi_2017_review,Feng_2017_review,Konotop_2016_review}, as well as, the existing experiments in quantum regime which do not feature a gain medium \cite{Xiao2017,Klauck2019, Li2019,Non_hermitian_AB_ring_expt_2020,Wu2019,
Naghiloo2019,EP_superconducting_circuits_expt_2019,
Anti_PT_quantum_experiment_2020,
Non_hermitian_conductance_expt_2019,Passive_PT_critical_phenomena_expt_2019}. 

\section{Comparison with local Lindblad results}
\label{Sec:local Lindblad}
All the results above are obtained from a complete microscopic Hamiltonian modelling of our set-up via an equation of motion approach. A much more common way of modelling gain-loss systems in quantum optics is via phenomenologically writing down Lindblad quantum master equations with local creation and annihilation operators of the cavities as Lindblad operators. To highlight the importance of our microscopic derivation, in this section we compare results from our completely microscopic equation of motion approach with two such local  Lindblad equations. We show that such local Lindblad equations can capture some qualitative features correctly, while missing some other qualitative aspects and predicting different results quantitatively.

\begin{figure}[!t]
\includegraphics[width=\columnwidth]{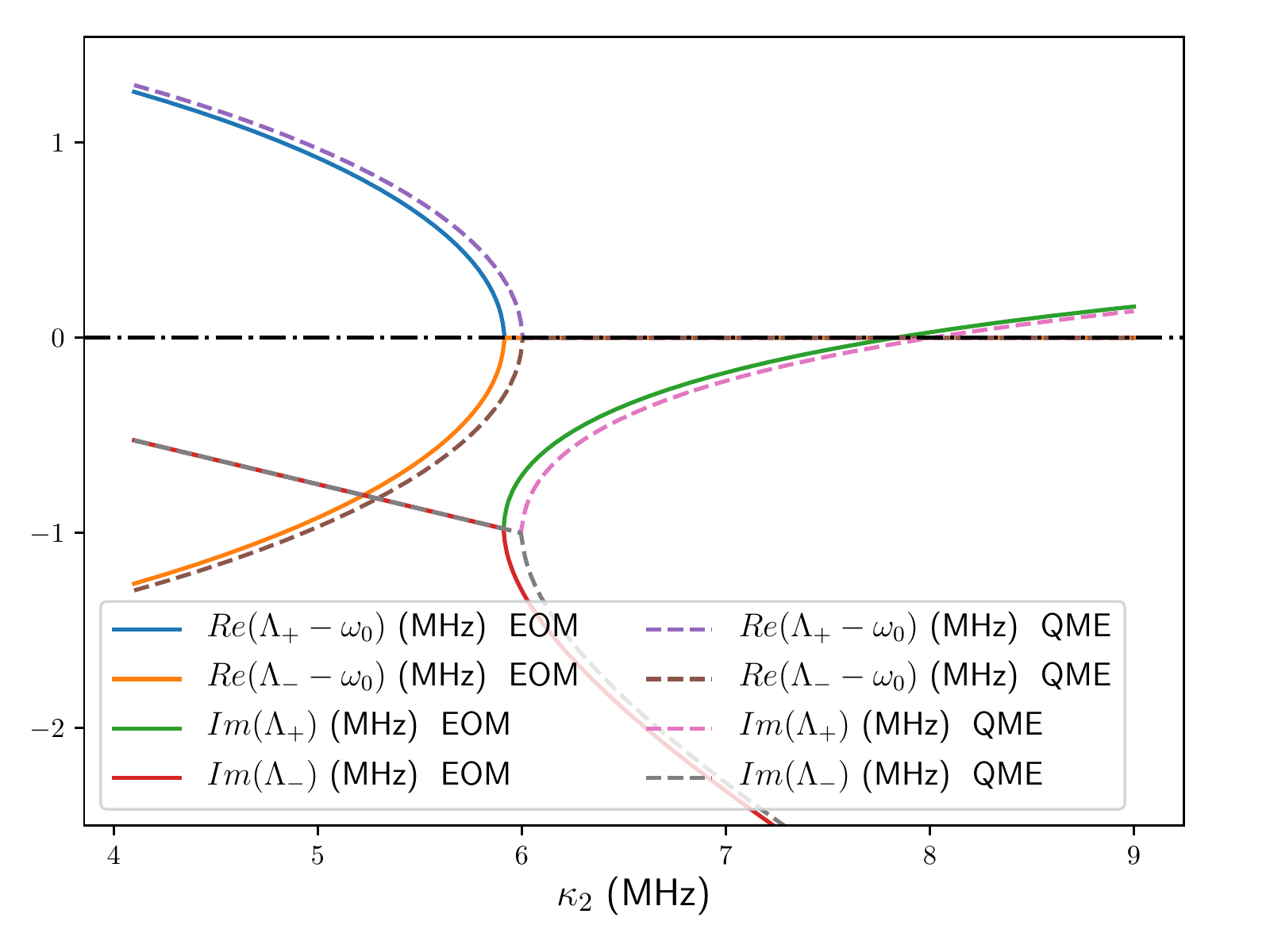} 
\caption{(color online) The real and imaginary parts of the eigenvalues of effective Hamiltonian $\mathbf{H}_{\rm eff}$ (Eq.(\ref{def_H_eff})) obtained by equations of motion approach (EOM) and those from the effective Hamiltonian $\mathbf{H}_{\rm eff}^{(2)}$ (Eq.(\ref{def_H_eff2})) obtained from local Lindblad quantum master equation (QME) are plotted as a function of the loss at the right cavity $\kappa_2$. Parameters are same as in Fig.~\ref{fig:PT_transition_kappa}.}
\label{fig:EP_kappa_QME_EOM} 
\end{figure}

The first local Lindblad equation that we consider can be microscopically derived for our set-up (see Appendix.~\ref{local Lindblad derivation}) using Born-Markov approximation, the conditions on energy scales in Eqs.(\ref{conditions_energy_scales_temp}), (\ref{conditions_energy_scales0}), (\ref{conditions_energy_scales}), along with the following approximation,
\begin{align}
\label{LLQME1_approx}
\hat{b}_\ell(t) \simeq e^{-i\omega_0 t} \hat{b}_\ell(0) + O(\lambda), 
\end{align}   
which is consistent with $\lambda\ll\omega_0$. This Lindblad equation is given by
\begin{align}
\label{LLQME1}
\frac{\partial\rho}{\partial t} &= i[\rho,\hat{\mathcal{H}}_C]+\frac{2g^2 \langle \hat{N}_1 \rangle_{ss}}{\Gamma} \left( \hat{b}_1^\dagger\rho\hat{b}_1 - \frac{1}{2}\{\hat{b}_1\hat{b}_1^\dagger,\rho\}\right) \nonumber \\
& + \left(\frac{2g^2 \langle \hat{N}_2 \rangle_{ss}}{\Gamma} +\kappa_1\right) \left( \hat{b}_1\rho\hat{b}_1^\dagger - \frac{1}{2}\{\hat{b}_1^\dagger\hat{b}_1,\rho\}\right)  \\
& + \kappa_2 \left( \hat{b}_2\rho\hat{b}_2^\dagger - \frac{1}{2}\{\hat{b}_2^\dagger \hat{b}_2,\rho\}\right),\nonumber
\end{align}
where $\hat{\mathcal{H}}_C$ is the Hamiltonian of the two coupled cavities (see Eq.(\ref{eq:DQDC})), and $\{\hat{P},\hat{Q}\}=\hat{P}\hat{Q}+\hat{Q}\hat{P}$ is the anti-commutator.
Multiplying the above equation by the corresponding operators and taking trace, the following equations for the expectation values of the cavity field operators and the cavity bilinears can be obtained,
\begin{align}
\label{LLQME1_ops1}
&i\frac{d}{dt}\left(
\begin{array}{c}
\langle \hat{b}_1 \rangle\\
\langle \hat{b}_2 \rangle\\
\end{array}
\right) = \mathbf{H}_{\mathrm{eff}}^{(2)}\left(
\begin{array}{c}
\langle \hat{b}_1 \rangle\\
\langle \hat{b}_2 \rangle\\
\end{array}
\right),  \\
\label{LLQME1_ops2}
& \frac{d \mathbf{C}}{dt} = i \mathbf{H}_{\mathrm{eff}}^{(2)^\dagger} \mathbf{C} - i \mathbf{C} \mathbf{H}_{\mathrm{eff}}^{(2)} + \frac{g^2 \langle \hat{N}_1 \rangle_{ss}}{\Gamma} (\mathbbm{1}_2+\sigma_z), 
\end{align}
where,
\begin{align}
\label{def_H_eff2}
&\mathbf{H}_{\mathrm{eff}}^{(2)}=\left[
\begin{array}{cc}
\omega_0-i\frac{\kappa_1}{2}+i\Gamma\delta & \lambda\\
\lambda & \omega_0-i\frac{\kappa_2}{2}\\
\end{array}\right],
\end{align} 
and the matrix $\mathbf{C}$ is as defined in Eq.(\ref{def_C}), and $\delta$ is as defined in Eq.(\ref{def_delta}).
Comparing above equation with Eq.(\ref{def_H_eff}), we see that the asymmetric off-diagonal terms are not captured by this local Lindblad approach. This will cause exceptional points of $\mathbf{H}_{\mathrm{eff}}^{(2)}$ to be slightly shifted from those of $\mathbf{H}_{\mathrm{eff}}$. This will, in turn, also shift other important values, like the threshold value of $\kappa_2$ for loss-induced lasing. But, for our choice of parameters, $\delta \ll 1$, so this shift is small. This is shown in Fig.~\ref{fig:EP_kappa_QME_EOM}.  Further, the formal solution of Eq.(\ref{LLQME1_ops2}) is given by
\begin{align}
\label{LLQME1_ops2_soln}
\mathbf{C}(t) &=  e^{i\mathbf{H}_{\mathrm{eff}}^{(2)^\dagger}t}\mathbf{C}(0)e^{-i\mathbf{H}_{\mathrm{eff}}^{(2)}t} \nonumber \\
&+\frac{g^2 \langle \hat{N}_1 \rangle_{ss}}{\Gamma} \int_0^t dt^\prime e^{i\mathbf{H}_{\mathrm{eff}}^{(2)^\dagger}t^\prime} (\mathbbm{1}_2+\sigma_z) e^{-i\mathbf{H}_{\mathrm{eff}}^{(2)}t^\prime}.  
\end{align}
Comparing this with Eq.(\ref{C_dynamics}), we see that  the effect of quantum fluctuations from the gain medium is not exactly the same. Nevertheless, if $\delta \ll 1$ and $\lambda\ll \omega_0$, since both approaches are microscopic, one can expect to get an approximate quantitative agreement between results from Eq.(\ref{LLQME1_ops2}) and the more accurate results from Eq.(\ref{formal_solution}). We will like to stress that obtaining Eq.(\ref{LLQME1}) required one more approximation, Eq.(\ref{LLQME1_approx}), which was not required in obtaining the results by our microscopic equations of motion approach. A more accurate quantum master equation can be derived without making this approximation, but that will not be of local Lindblad form \cite{Lindblad_validity_1,Lindblad_validity_2,Lindblad_validity_3}. 

For a given value of $2\Gamma\delta-\kappa_1>0$, $\mathbf{H}_{\mathrm{eff}}^{(2)}$ describes an effective Hamiltonian of the coupled cavities with gain in the first site and loss in the second site. In many cases (for example, \cite{Wunner2014, Wunner2018, Quantum_PT_2020_3,Quantum_PT_2020_1}), when the objective is to model such an effective Hamiltonian with a given amount of gain and loss, a completely phenomenological Lindblad equation is written down as follows,
\begin{align}
\label{LLQME2}
\frac{\partial\rho}{\partial t} &= i[\rho,\hat{\mathcal{H}}_C]+ \left(2\Gamma \delta-\kappa_1\right) \left( \hat{b}_1^\dagger\rho\hat{b}_1 - \frac{1}{2}\{\hat{b}_1\hat{b}_1^\dagger,\rho\}\right) \nonumber \\
& + \kappa_2 \left( \hat{b}_2\rho\hat{b}_2^\dagger - \frac{1}{2}\{\hat{b}_2^\dagger \hat{b}_2,\rho\}\right).
\end{align}
This equation cannot be microscopically derived for our set-up, unless the DQD is completely population inverted, i.e, $\langle \hat{N}_1 \rangle_{ss}=1$, $\langle \hat{N}_2 \rangle_{ss}=0$, and the first site has no intrinsic loss, i.e, $\kappa_1=0$. For given amount of loss in second cavity, $\kappa_2$, this condition need not be satisfied for obtaining $\mathcal{PT}$-symmetry in the effective Hamiltonian. Further, while phenomenological local Lindblad equation is designed such that the equations for the expectation values of the cavity field operators are exactly the same as in Eq.(\ref{LLQME1_ops1}), that for the cavity bilinears is given by
\begin{align}
\label{LLQME2_ops}
&\frac{d \mathbf{C}}{dt} = i \mathbf{H}_{\mathrm{eff}}^{(2)^\dagger} \mathbf{C} - i \mathbf{C} \mathbf{H}_{\mathrm{eff}}^{(2)} + \left(\Gamma \delta-\frac{\kappa_1}{2}\right)~ (\mathbbm{1}_2+\sigma_z), \\
&\Rightarrow\mathbf{C}(t) =  e^{i\mathbf{H}_{\mathrm{eff}}^{(2)^\dagger}t}\mathbf{C}(0)e^{-i\mathbf{H}_{\mathrm{eff}}^{(2)}t} \nonumber \\
& \hspace{40pt} +\left(\Gamma \delta-\frac{\kappa_1}{2}\right) \int_0^t dt^\prime e^{i\mathbf{H}_{\mathrm{eff}}^{(2)^\dagger}t^\prime} (\mathbbm{1}_2+\sigma_z) e^{-i\mathbf{H}_{\mathrm{eff}}^{(2)}t^\prime} \nonumber
\end{align}
Clearly, the strength of quantum fluctuations of the gain medium, embodied in the second term in above equation, is completely different from what is seen in both the microscopic derivations. Thus, the purely  phenomenological approach will severely underestimate the strength of quantum fluctuations, unless under very special conditions. This is despite the fact that the complex quadratures $\langle \hat{b}_1(t) \rangle$, $\langle \hat{b}_2(t) \rangle$,  from all three approaches will be the almost same if $\delta\ll 1$.

The formal solutions for $\mathbf{C}(t)$ in Eqs.(\ref{LLQME1_ops2_soln} and (\ref{LLQME2_ops}) are of the same form. For any equation of that form, it can be checked by direct calculation that if the eigenvalues of $\mathbf{H}_{\mathrm{eff}}^{(2)}$ are real ($\mathcal{PT}$-symmetric phase), the inhomogeneous part will lead to a linear divergence with time in $\mathbf{C}$. It can also be checked that the inhomogeneous part will lead to a $t^3$ divergence when $\mathbf{H}_{\mathrm{eff}}^{(2)}$ is at the exceptional point, and will lead to exponential divergence if one of the eigenvalues of $\mathbf{H}_{\mathrm{eff}}^{(2)}$ have a positive imaginary part ($\mathcal{PT}$-broken phase). This behavior with time is exactly the same as described in Sec.~\ref{Sec: Balanced gain-loss} via our equation of motion approach.  Thus, even though both the Lindblad equations do not capture the location of the exceptional point and the strength of quantum fluctuations accurately, both of them capture the qualitative behavior with time correctly in either side of the $\mathcal{PT}$-transition, as well as at the exceptional point. Features of dissipative exceptional points, like loss induced increase in average photon number in absence of any coherent drive, will also be captured qualitatively by both Lindblad equations, though not quantitatively. This shows that these effects of quantum fluctuations are actually more general than the set-up we have considered, and will generically hold.  However, from our microscopic modelling, we know that all such results from linearized descriptions are valid only when $g \sqrt{n_\mathrm{photons}} \leq O(\Gamma)$ can be satisfied. Beyond this regime, non-linear effects need to be considered, and none of the linearized descriptions, including our linearized microscopic equation of motions approach, hold. This fact would not be clear in a purely phenomenological approach.

\begin{figure}[!t]
\includegraphics[width=\columnwidth]{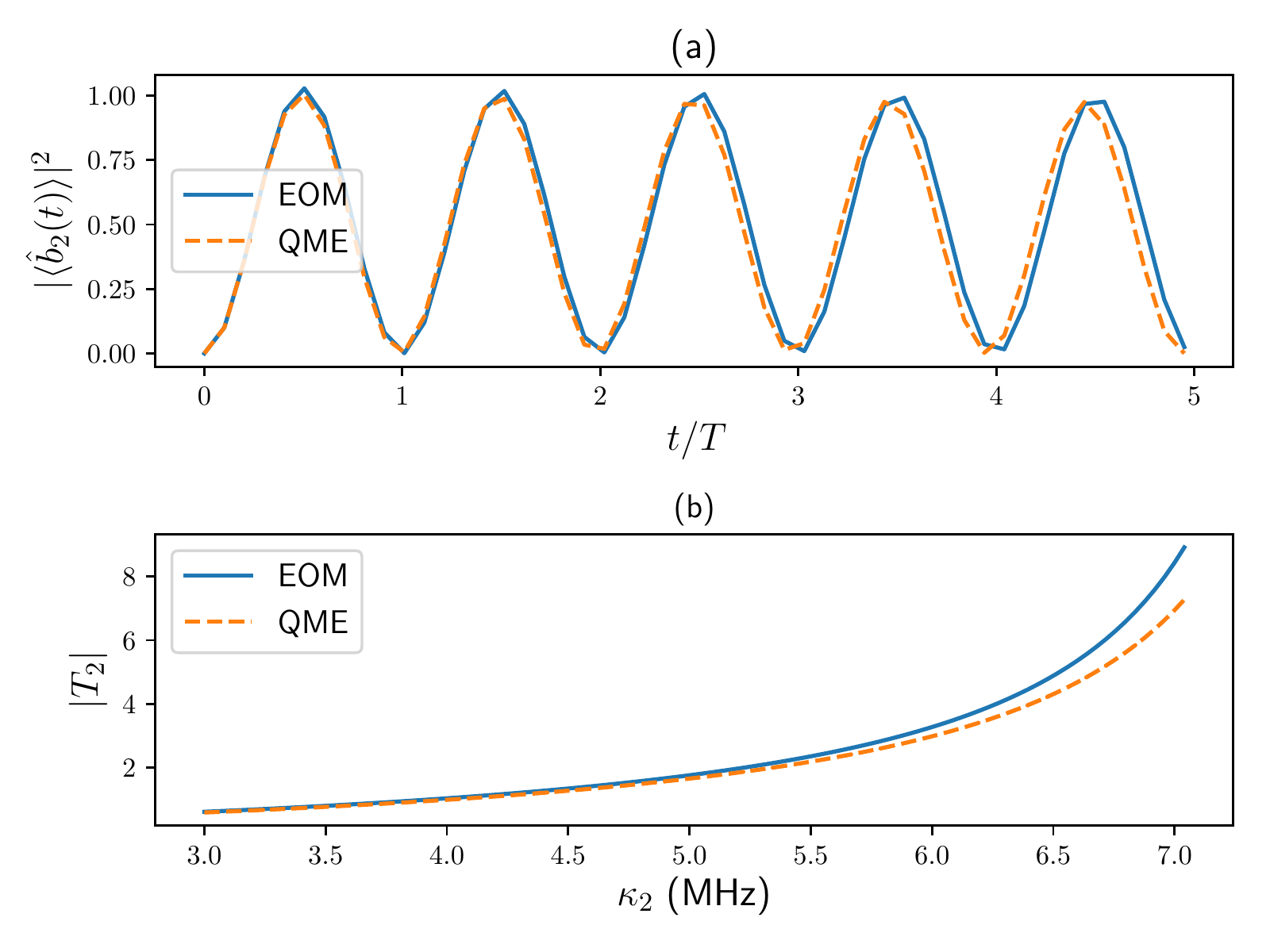} 
\caption{(color online) (a) The plot shows $\mathcal{PT}$-symmetric dynamics of $|\langle \hat{b}_2(t) \rangle|^2$ as obtained from equations of motion approach (EOM), Eq.~\ref{eq:eom}, and as from the local Lindblad approach (QME), Eq.~\ref{LLQME1_ops1}. The initial condition, the parameters and the period of oscillation $T$ are same as in Fig.~\ref{fig:PT_dynamics}. (b) The plot shows the amplitude of transmitted signal at the right cavity for an input-output experiment with a weak coherent drive also at the right cavity, as a function of the loss at the right cavity $\kappa_2$, at resonance $\omega_d=\omega_0$. The results obtained both from the equations of motion approach (EOM) and local Lindblad approaches (QME) are shown. The parameters are exactly same as in Fig.~\ref{fig:PT_transition_kappa}.}
\label{fig:b2_QME_EOM} 
\end{figure}

Now we numerically check the results obtained from both the Lindblad quantum master equations against the equaiton of motion approach. In Fig.~\ref{fig:b2_QME_EOM}, we compare properties which depend on the expectation values of the complex quadratures. In Fig.~\ref{fig:b2_QME_EOM}(a) we compare $\mathcal{PT}$-symmetric dynamics of $|\langle \hat{b}_2(t) \rangle|^2$ starting from a coherent state in the left cavity, and an empty state in the right cavity. Since our choice of $\lambda=10$ MHz, $\omega_0=8$ GHz, (same as in Fig.~\ref{fig:PT_dynamics}) satisfies $\lambda \ll \omega_0$, the results obtained from the local Lindblad equaiton match quite well with those from the microscopic equations of motion. However, the small mismatch grows with increase in time. In Fig.~\ref{fig:b2_QME_EOM}(b) we compare the loss-induced enhancement of amplification for a weak coherent drive at resonance $\omega_d=\omega_0$ as obtained from both approaches. The results agree well for smaller values of $\kappa_2$, while the difference between them increases with increase in $\kappa_2$. In both these cases, the difference between the results stem from the presence of asymmetric hopping in $\mathbf{H}_{\mathrm{eff}}$ (Eq.(\ref{def_H_eff})), which is absent in $\mathbf{H}_{\mathrm{eff}}^{(2)}$ (Eq.(\ref{def_H_eff2})). We remind the reader that both the microscopically derived local Lindblad equation, Eq.(\ref{LLQME1}),  and the purely phenomenologically written local Lindblad equation, Eq.(\ref{LLQME2}), give exactly same results for these quantities, governed by $\mathbf{H}_{\rm eff}^{(2)}$ (Eq.(\ref{def_H_eff2})). 

\begin{figure}[!t]
\includegraphics[width=\columnwidth]{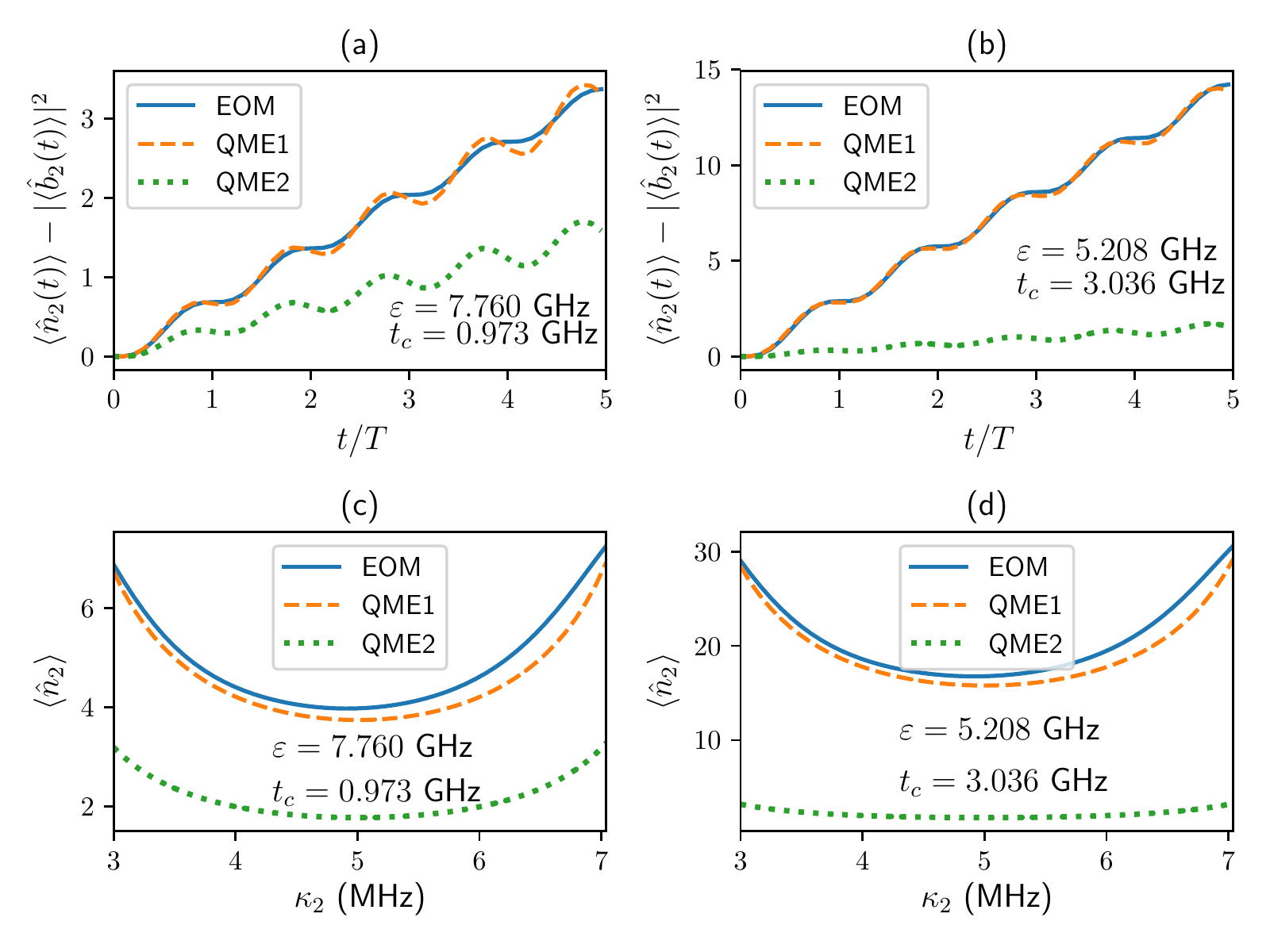} 
\caption{(color online) The figure compares the effect of quantum fluctuations of the gain medium as obtained from the microscopic equations of motion approach (EOM), Eq.(\ref{eq:eom},) with those from the microscopically derived local Lindblad approach (QME1), Eq.(\ref{LLQME1}), and from the purely phenomenological local Linblad approach (QME2), Eq.(\ref{LLQME2}). (a) Quantum fluctuations in the $\mathcal{PT}$-symmetric dynamics of the complex quadrature of the right cavity as a function of time, for parameters same as in Fig.~\ref{fig:PT_dynamics}. (b) Exactly same as in (a), except for different values of $\varepsilon$ and $t_c$, corresponding to the second balanced-gain-loss condition (see Fig.~\ref{fig:N_ss_sin}). (c) The average photon number of the right cavity in steady state as a function of the loss of the right cavity $\kappa_2$, in absence of any coherent drive. The parameters are as in Fig.~\ref{fig:PT_transition_kappa}. (d) Exactly same as in (c), except for different values of $\varepsilon$ and $t_c$.  }
\label{fig:occ2_QME_EOM} 
\end{figure}

In Fig.~\ref{fig:occ2_QME_EOM}, we compare the quantum fluctuations in the complex quadrature of the right cavity as obtained from the three approaches. Here, we look at both of the two choices of $\varepsilon$ and $t_c$ that lead to a balanced-gain-loss condition (see Fig.~\ref{fig:N_ss_sin}). As mentioned before in Sec.~\ref{Sec: Effective PT}, these two conditions have widely different values of population inversion $\Delta N_{ss}$, given in Eq.(\ref{population_inversion}), leads to the same $\mathbf{H}_{\rm eff}$. However, the strength of quantum fluctuations of the gain medium is different in these two cases.  This is shown in Fig.~\ref{fig:occ2_QME_EOM}(a) and (b) where the $\mathcal{PT}$-symmetric dynamics of $\langle\hat{n}_2(t)\rangle-|\langle\hat{b}_2(t)\rangle|^2$ is shown for both cases from all three approaches. It is clear that the strength of quantum fluctuations is different in the two cases. Since our choice of parameters (same as in Fig.~\ref{fig:PT_dynamics}, Fig~\ref{fig:b2_QME_EOM}) satisfies $\lambda \ll \omega_0$, the results from the microscopic Lindblad equation Eq.(\ref{LLQME1}) match quite well with those from the microscopic equation of motion approach. But, since the left cavity with gain also has an intrinsic loss, the results from the purely phenomenological approach Eq.(\ref{LLQME2}) do not agree with the other two approaches. The discrepancy between the purely phenomenological approach and the two microscopic approaches is even more for the second choice of $\varepsilon$ and $t_c$. This is because, this choice corresponds to small population inversion.  However, all three approaches show the linear divergence with time for both values of $\varepsilon$ and $t_c$.  Note that, dynamics of
$|\langle\hat{b}_2(t)\rangle|^2$ for both choices of $\varepsilon$ and $t_c$ will be same and as given in Fig.~\ref{fig:b2_QME_EOM}(a). Fig.~\ref{fig:occ2_QME_EOM}(c) and (d) shows the non-monotonic behavior of the steady state average photon number of the right cavity $\langle\hat{n}_2\rangle$ as a function of $\kappa_2$ in absence of any coherent drive for the two different choices of $\varepsilon$ and $t_c$. In this case, also, the purely phenomenological approach severely underestimates $\langle\hat{n}_2\rangle$, which depends on the quantum fluctuations of the gain medium, for the same reasons as above. The microscopic Lindblad equation also does not agree well with microscopic equation of motion results, especially at larger values of $\kappa_2$, although $\lambda=20$~MHz is still satisfies $\lambda\ll \omega_0$. It fails to capture the slightly steeper rise in $\langle\hat{n}_2\rangle$ with increase in $\kappa_2$. However, all three approaches capture the qualitative feature of non-monotonic behavior of average photon number of second cavity with increase in loss at the second cavity.   


The analysis in this section highlights the importance of complete microscopic modelling of experimental set-ups for exploring non-Hermitian physics in quantum regime. In most experimental set-ups with coupled gain and loss cavities, the cavity with effective gain will have also have some loss rate, which has to be compensated by the gain rate to achieve the effective gain. The results in this section show that a phenomenological quantum master equation of the form Eq.(\ref{LLQME2}) with a gain term in Lindblad form in one cavity and a loss term in Lindblad form in the other cavity is inadequate to accurately describe any such system. Even when the gain and loss parameters are chosen to give the expectation values of complex quadratures accurately, such phenomenological Lindblad equations will always severely underestimate the quantum fluctuations in the complex quadratures. A minimal phenomenological model to describe such a system in terms of a local Lindblad equation has to have the form of Eq.(\ref{LLQME1}), with three parameters, viz. the gain rate of one cavity, and the loss rates of both the cavities. As long as the coupling between the cavities is small, such a Lindblad equation can reasonably accurately capture both the expectation values and the quantum fluctuations in the complex quadratures of the cavities. However, even such a Lindblad equation will not capture the effect of the asymmetric hopping, and will underestimate the loss-induced increase in average photon number.  Nevertheless, some universal qualitative features will be captured correctly by all approaches.

\section{Summary and outlook}
\label{Sec: Conclusions}

In this paper, we have shown that a set-up of two coupled cQED cavities with a DQD in one of them can be used to explore the physics of $\mathcal{PT}$-symmetry breaking transition, including the effect of quantum fluctuations. As has already been established in several experiments~\cite{photon_emission_DQD_cQED,DQD_cQED_floquet_gain,DQD_micromaser,
on_chip_DQD_light_source,
probing_electron_phonon_DQD_cQED_1}, a DQD in a cavity can act as a highly tunable gain medium for the cavity. In this work, under a reasonable approximation on various energy scales consistent with state-of-the-art experiments, we have microscopically derived an effective non-Hermitian Hamiltonian, along with the quantum noise term, which govern the dynamics of bosonic complex quadratures that describe the cavity. This derivation requires a linearization of the non-linear interaction between the DQD and the cavity and only holds if the number of photons in the cavity is not too large. We have also kept track of the quantum fluctuations coming from the presence of the DQD. We have thus obtained the signatures of the $\mathcal{PT}$ transition in the expectation values of the field operators (complex quadratures), and signatures of quantum fluctuations in the expectation values of field-operator bilinears (cavity photon numbers and intercavity current). In both $\mathcal{PT}$-symmetric phase and at the exceptional point, we have found that quantum fluctuations lead to a linear-in-time factor over the classical predictions for photon numbers and photon current. These signatures are then generalized to dissipative cavities that reach a steady state. Here, the passive $\mathcal{PT}$ transition is marked by the bifurcation of the decay rates for the two dissipative eigenmodes of the system. We have shown that this can be accurately tracked in a particular input-output experiment. We have further shown that loss-induced enhancement of amplification and lasing can be observed in the set-up as a consequence of a transition across an exceptional point.  We have identified loss-induced increase of average photon number in absence of any coherent drive as a particular property of a quantum non-Hermitian system featuring a gain medium. Finally, we have highlighted the importance of our microscopic derivation, by comparing our results with two different local Lindblad equations. Our analysis has also pointed at some model independent universal features of quantum fluctuations in gain-loss systems as well as at the minimal Lindblad equation required to describe a realistic quantum $\mathcal{PT}$-symmetric system reasonably accurately in a parameter regime.  

This work has several important consequences.  The set-up of a DQD in a cQED cavity is already state-of-the-art \cite{wallraff2019_2,wallraff2019_1, photon_emission_DQD_cQED,DQD_micromaser,Phonon_assisted_gain_DQD_cQED,
DQD_cQED_floquet_gain,on_chip_DQD_light_source,
probing_electron_phonon_DQD_cQED_1,probing_electron_phonon_DQD_cQED_2,
Kontos_DQD_cQED_review_2} and connecting several cavities is also a standard technique in cQED experiments \cite{cQED_review}.  We have shown that the resulting set-up provides a highly tunable system where the exotic physics of exceptional point degeneracies in the quantum regime can be explored, along with a systematic study of quantum fluctuations that accompany any gain medium~\cite{Caves1982,Scheel2018}. This set-up is potentially scalable \citep{72_resonators_cQED,12_dots}. Moreover, since a completely controllable gain can be introduced in an arbitrary cavity simply with a DQD in that cavity, our work presents the possibility of creating non-Hermitian bosonic systems with arbitrarily distributed gain and loss in the quantum regime. This is of particular importance for extending the forefronts of non-Hermitian topological systems~\cite{Zhen2015,Xu2016,Harter2016,Harari2018,Gong2018,Kawabata2019,
FoaTorres2019,Longhi_topological_quasicrystals_2019, Bandres2018,Topological_PT_2020,
Topological_PT_circuit_expt_2020,Non_Hermitian_topology_expt_2019,Xiao2017} into the quantum domain, as well as their possible applications in quantum technology \cite{Quantum_noise_EP_theory_2020,Quantum_sensing_single_qubit_theory_2020}. 

It is important to emphasize that a rigorous description showing the emergence of an effective parity-time symmetric non-Hermitian Hamiltonian starting from a microscopic Hermitian Hamiltonian model of open quantum system has been missing so far. Most models \cite{Ozdemir_2019_review,El-Ganainy_2018_review,Longhi_2017_review,Feng_2017_review,Konotop_2016_review} start with a non-Hermitian Hamiltonian (containing complex frequencies) that is applicable only in the classical limit where the number of energy quanta is much larger than one. 
For dissipative open quantum systems, one can write down a
Lindblad equation from which one
can get a non-Hermitian Hamiltonian by ignoring quantum jumps \cite{Avila2020,Quantum_PT_2020,Klauck2019,Naghiloo2019,
Rotter_news_and_views_2019,Graefe_news_and_views_2019,Yamamoto2019}. This approach has been generalized to gain-loss systems, with and without neglecting the quantum jumps, by writing down Lindblad terms for gain and loss phenomenologically \cite{Wunner2014, Wunner2018,Quantum_PT_2020_1,Quantum_PT_2020_3,
Rabl2019_power_grid}. More microscopic approaches, which involve modelling the gain medium, also start from phenomenological local Lindblad descriptions of various losses involved in the set-up \cite{Rabl2019_spin_lattice,PT_symmetric_cQED}. Our work starting from a completely microscopic Hermitian Hamiltonian of a state-of-the-art experimental set-up is therefore a major step forward from all such phenomenological constructions. Such a microscopic description in terms of operator equations of motion also facilitates calculation of two-time correlations, which is often a challenge in the master equation approaches. Further, it opens the possibility of exploring the quantum thermodynamics of effective $\mathcal{PT}$-symmetric systems theoretically as well as experimentally.  Though both technologically and fundamentally important, this is a completely uncharted territory at present, and impossible to explore with previous phenomenological techniques. We will investigate these directions in future works. Although, our work was based on using a voltage-biased DQD as a gain medium, our findings and methods can be adapted to other potential quantum non-Hermitian systems with a gain medium governed by population inversion.

\section{Acknowledgement}

We would like to thank Jason Petta for useful discussions. M. K.
gratefully acknowledges the Ramanujan Fellowship SB/S2/RJN-114/2016
from the Science and Engineering Research Board (SERB), Department of
Science and Technology, Government of India. M. K. also acknowledges
support from the Early Career Research Award, ECR/2018/002085  from
the Science and Engineering Research Board (SERB), Department of
Science and Technology, Government of India. MK would like to
acknowledge support from the project 6004-1 of the Indo-French Centre
for the Promotion of Advanced Research (IFCPAR). M. K. acknowledges
support from the Matrics Grant, MTR/2019/001101 from the Science and
Engineering Research Board (SERB), Department of Science and
Technology, Government of India. We are grateful for the hospitality
of International Centre for Theoretical Sciences (ICTS) - Tata
Institute of Fundamental Research, Bangalore during the program on
"Non-Hermitian Physics - PHHQP XVIII" (ICTS/nhp2018/06) where this
work was initiated. A.P. acknowledges funding from the
European Research Council (ERC) under the European
Unions Horizon 2020 research and innovation program (Grant
Agreement No. 758403). YJ acknowledges funding from NSF grant DMR-1054020.

\appendix


\section{Derivation of the effective non-Hermitian Hamiltonian for the cavity unit}\label{Derivation}

\subsection{Isolated cavities}
We first solve for the dynamics of the two cavities in isolation, i.e, without coupling to bosonic baths, or the DQD. To this end, we note that, the Hamiltonian governing the two cavities can be written as
\begin{align}
\label{def_H_S}
&\hat{\mathcal{H}}^{b}_{S} = \left(\hat{b}^{\dagger}_1~~\hat{b}^{\dagger}_2 \right) \mathbf{H}_S\left(
\begin{array}{c}
\hat{b}_1\\
\hat{b}_2\\
\end{array}
\right),\nonumber \\
&\mathbf{H}_S = \omega_0 \mathbb{I}+ \lambda \sigma_x, 
\end{align}
where $\mathbb{I}$ is the $2$x$2$ identity matrix and $\sigma_x$ is the usual Pauli matrix. The matrix $\mathbf{H}_S$ is digonalized by the following orthogonal matrix,
\begin{align}
\label{cavity_normal_modes}
&{\Phi^{b}}^T \mathbf{H}_S {\Phi^{b}}= \left(
\begin{array}{cc}
\omega_1 & 0\\
0 & \omega_2\\
\end{array}
\right),~~\Phi^b = \frac{1}{\sqrt{2}}\left(
\begin{array}{cc}
1 & -1\\
1 & 1\\
\end{array}
\right), \nonumber \\
&\omega_1= \omega_0+g,~~\omega_2=\omega_0-g.
\end{align}
The isolated dynamics of the two cavities is then given by
\begin{align}
\hat{b}_\ell(t_0) = \sum_{p,\alpha=1}^2 \Phi_{\alpha \ell}^b \Phi_{\alpha p}^b e^{i\omega_{\alpha} (t-t_0)} \hat{b}_p(t).
\end{align}
Now, we go ahead to find effective evolution equations for the cavity operators in presence of bosonic baths and the DQD, by integrating out everything other than the cavities.

\subsection{Setting the relative strength of various energy scales}
In order to integrate out everything other than the cavities, we need to keep track of the relative strengths of various energy scales. To keep track of relative strengths of various energy scales, we introduce a dimensionless small parameter $\epsilon \ll 1$ and consider
\begin{align}
\kappa_{s\ell} \rightarrow \epsilon^2 \kappa_{s\ell},~\Gamma_{sl}\rightarrow \epsilon \Gamma_{sl},~\lambda_s^{ph}\rightarrow \epsilon \lambda_s^{ph}, ~g_0 \rightarrow \epsilon^2 g_0. 
\end{align}
Our goal is to integrate out the bosonic baths as well as the DQD unit to obtain effective equations of motion for the two coupled cavities up to $O(\epsilon^4)$. In the following, we do this part by part.

\subsection{Integrating out bosonic baths}
The cavities are assumed to be coupled to the bosonic baths as well as to the DQD unit at time $t=0$.  The state of the entire set-up unit at time $t=0$ is taken to be
\begin{align}
\rho_{tot}(0)=\rho^{DQD}(0)\frac{e^{-\beta\hat{\mathcal{H}}^{(1)}_{B}}}{Z_1^b}  \rho_S^b \frac{e^{-\beta\hat{\mathcal{H}}^{(2)}_{B}}}{Z_2^b}, 
\end{align}
where $\beta$ is the inverse temperature of the baths, $\hat{\mathcal{H}}_{B}^{(\ell)}= \sum_{s=1}^{\infty}\Omega_{s\ell} \hat{B}_{s\ell}^\dagger \hat{B}_{s\ell}$ is the Hamiltonian of the bath attached to $\ell$th cavity, $Z_\ell^b$ is the partition function of the initial state of $\ell$th bath and $\rho^{DQD}(0)$ is the state of the DQD at time $t=0$.  The initial state of the two coupled cavities is $\rho_S^b$, which is arbitrary.

To integrate out the bath degrees of freedom, let us first write down the Heisenberg equations of motion for all the bosonic system operators,
\begin{align}
\label{boson_heisenberg}
&\frac{d\hat{b}_1}{dt} = -i\omega_0 \hat{b}_1 -i \lambda \hat{b}_2 -i \epsilon^2 \sum_{s=1}^\infty \kappa_{s 1} \hat{B}_{s1}-i\epsilon^2 g \Theta(t) \hat{A}_2^\dagger\hat{A}_1, \nonumber \\
&\frac{d\hat{b}_2}{dt} = -i\omega_0 \hat{b}_2 -i \lambda \hat{b}_1 -i \epsilon^2\sum_{s=1}^\infty \kappa_{s 2} \hat{B}_{s2}. 
\end{align}
The formal solution for the bosonic bath operators is given by
\begin{align}
\hat{B}_{s\ell}(t)=e^{-i\Omega_{s\ell} t}\hat{B}_{s\ell}(0)-i \epsilon^2 \kappa_{s \ell} \int_0^t dt^\prime e^{-i\Omega_{s\ell} (t-t^\prime)}\hat{b}_\ell(t^\prime).
\end{align}
Using these formal solutions in Eq.(\ref{boson_heisenberg}), we obtain,
\begin{align}
\label{boson_effective_exact}
&\frac{d\hat{b}_1}{dt} = -i\omega_0 \hat{b}_1 -i \lambda \hat{b}_2-i\epsilon^2 g \Theta(t) \hat{A}_2^\dagger\hat{A}_1 -i \epsilon^2 \hat{\xi}_1^b(t) \nonumber \\
&- \epsilon^4 \int_0^t dt^\prime \int_{0}^{\infty}\frac{d\omega}{2\pi}\mathfrak{J}_1 (\omega)e^{-i\omega (t-t^\prime)}\hat{b}_1(t^\prime)  , \nonumber \\
&\frac{d\hat{b}_2}{dt} = -i\omega_0 \hat{b}_2 -i \lambda \hat{b}_1 -i \epsilon^2 \hat{\xi}_2(t) \nonumber \\
&- \epsilon^4 \int_0^t dt^\prime \int_{0}^{\infty}\frac{d\omega}{2\pi}\mathfrak{J}_2 (\omega)e^{-i\omega (t-t^\prime)}\hat{b}_2(t^\prime), 
\end{align}
where
\begin{align}
\label{def_xib}
&\hat{\xi}_\ell^b(t) = \sum_{s=1}^\infty \kappa_{s \ell} e^{-i\Omega_{s\ell} t}\hat{B}_{s\ell}(0). 
\end{align}
Because of our chosen initial state, we get,
\begin{align}
\label{bath_noise_correlations}
& \langle \hat{\xi}_\ell^b(t) \rangle = 0, \nonumber \\
&\langle \hat{\xi}_\ell^{b\dagger}(t)\hat{\xi}_m^b(t^\prime)\rangle= \delta_{\ell m} \int_{0}^{\infty}\frac{d\omega}{2\pi}\mathfrak{J}_\ell (\omega)\mathfrak{n}^B(\omega)e^{i\omega(t-t^\prime)}, 
\end{align}
where 
\begin{align}
\mathfrak{n}^B(\omega) = \frac{1}{e^{\beta\omega}-1},
\end{align}
is the bose distribution.
Now, using the definitions in Eq.(\ref{cavity_normal_modes}), we note that,
\begin{align}
\label{b_perturb}
\hat{b}_\ell(t^\prime) = \sum_{p,\alpha=1}^2 \Phi_{\alpha \ell}^b \Phi_{\alpha p}^b e^{i\omega_{\alpha} (t-t^\prime)} \hat{b}_p(t) + O(\epsilon^2).
\end{align}
So we obtain up to $O(\epsilon^4)$,
\begin{align}
&\epsilon^4 \int_0^t dt^\prime \int_{0}^{\infty}\frac{d\omega}{2\pi}\mathfrak{J}_\ell (\omega)e^{-i\omega (t-t^\prime)}\hat{b}_\ell(t^\prime)\simeq \epsilon^4 \sum_{p=1}^2 v_{\ell p}^b(t) \hat{b}_p(t), \nonumber \\
&v_{\ell p}^b(t) = \sum_{\alpha=1}^2 \Phi_{\alpha \ell}^b  \Phi_{\alpha p}^b \int_{0}^t dt^\prime \int_{0}^{\infty} \frac{d\omega}{2\pi} \mathfrak{J}_\ell(\omega) e^{-i(\omega-\omega_\alpha)t^\prime}.
\end{align}
Let $\tau_{B_\ell}$ be the time in which $\int_{0}^{\infty} \frac{d\omega}{2\pi} \mathfrak{J}_\ell(\omega) e^{-i\omega t^\prime}$ decays to $O(\epsilon)$. Then, for $t\gg \tau_{B_\ell}$, we have,
\begin{align}
v_{\ell p}^b(t) \simeq v_{\ell p}^b(\infty)\equiv v_{\ell p}^b.
\end{align}
If $\mathfrak{J}_\ell(\omega)$ is such that it is nearly flat around $\omega_0$ on a scale much larger than $\lambda$, and is given by $\mathfrak{J}_\ell(\omega)\simeq \kappa_\ell$, then to a good approximation,  we can write, 
\begin{align}
\label{const_spec_approx}
v_{\ell p}^b \simeq \frac{\kappa_\ell}{2} \delta_{\ell p}.
\end{align}
So, finally, we obtain,
\begin{align}
\label{bosons_bath_integrated}
&\frac{d\hat{b}_1}{dt} = -i\omega_0 \hat{b}_1 -i \lambda \hat{b}_2-i\epsilon^2 g\Theta(t) \hat{A}_2^\dagger\hat{A}_1 -i \epsilon^2 \hat{\xi}_1^b(t) - \epsilon^4 \frac{\kappa_1}{2} \hat{b}_1  , \nonumber \\
&\frac{d\hat{b}_2}{dt} = -i\omega_0 \hat{b}_2 -i \lambda \hat{b}_1 -i \epsilon^2 \hat{\xi}_2^b(t)-
\epsilon^4 \frac{\kappa_2}{2} \hat{b}_2. 
\end{align}
In above effective equations of motion, we have the dissipative terms coming from the bosonic baths. If there was no boson-fermion coupling, i.e, $g=0$, we would have had time evolution by an effective non-Hermitian dissipative Hamiltonian. However, there would be an inhomogenous part to the equation coming from the `noise' terms, $\hat{\xi}_\ell^b(t)$ associated with the corresponding dissipations. The relavant noise correlations are given in Eq.(\ref{bath_noise_correlations}). The `noise' and dissipation coming from the baths satisfy the fluctuation-dissipation relations and embody quantum, as well as thermal, fluctuations. The connection to the DQD unit is coming from the term $i\epsilon^2 g \hat{A}_2^\dagger\hat{A}_1$ in above equation. Next we will integrate out the DQD unit.

\subsection{Integrating out fermions}
The DQD is assumed to be connected to the fermionic leads and the phononic substrate at time $t=t_0$, with $t_0\ll 0$. The state of the DQD unit at time $t=t_0$ is taken to be
\begin{align}
&\rho^{DQD}_{tot}(t_0)= \\
&\frac{e^{-\beta(\hat{\mathcal{H}}^{(1)}_{L}-\mu_1\hat{\mathcal{N}}^{(1)}_{L})}}{Z_1^f} \rho_{DQD}(t_0) \frac{e^{-\beta(\hat{\mathcal{H}}^{(2)}_{L}-\mu_2\hat{\mathcal{N}}^{(2)}_{L})}}{Z_2^f} \frac{e^{-\beta\hat{\mathcal{H}}_{ph}}}{Z_{ph}} \nonumber,
\end{align}
where $\hat{\mathcal{H}}^{(1)}_{L}=\sum_{s=1}^{\infty} \mathcal{E}_{sl}\hat{a}_{s\ell}^{\dagger}\hat{a}_{s\ell}$ is the Hamiltonian of the fermionic lead attached to the $\ell$th site of the DQD, $\hat{\mathcal{N}}^{(1)}_{L}=\sum_{s=1}^{\infty} \hat{a}_{s\ell}^{\dagger}\hat{a}_{s\ell}$ is the total number operator of the lead, $Z_\ell^f$, $Z_{ph}$ are corresponding partition functions, $\beta$, $\mu_1$ and $\mu_2$ are the corresponding inverse temperature and the chemical potentials. The initial state of the DQD is $\rho_{DQD}$, which is arbitrary.

Following exactly similar steps as for the bosons, the effective equation of motion for $\hat{A}_2^\dagger\hat{A}_1$, after integrating out the fermionic baths, can be written up  as
\begin{align}
&\frac{d (\hat{A}_2^\dagger\hat{A}_1)}{dt}= -i\omega_q\hat{A}_2^\dagger\hat{A}_1-\epsilon^2 \Gamma \hat{A}_2^\dagger\hat{A}_1  -i\epsilon^2 g \Theta(t) (\hat{N}_1-\hat{N}_2) \hat{b}_1 \nonumber \\
&-\frac{i\epsilon}{\omega_q}\left[2\varepsilon \hat{A}_2^\dagger\hat{A}_1-2t_c(\hat{N}_1-\hat{N}_2)\right]\sum_{s=1}^{\infty}
\lambda_{s}^{ph} \left(\hat{B}_{s}^{ph\dagger} + \hat{B}_{s}^{ph}\right) \nonumber \\
& - i\epsilon(\hat{A}_2^\dagger \hat{\xi}_1^f - \hat{\xi}_2^{f\dagger}\hat{A}_1),
\end{align}
where,
\begin{align}
&\hat{\xi}_\ell^f(t) = \sum_{s=1}^\infty \Gamma_{s \ell}e^{-i\mathcal{E}_{s\ell} t}\hat{a}_{s\ell}(0). 
\end{align}
The formal solution of this equation can be written as
\begin{align}
\label{A2A1_exact_all}
&\hat{A}_2^\dagger(t)\hat{A}_1(t)=    e^{-(i\omega_q+\epsilon^2\Gamma)t} \hat{A}_2^\dagger(t_0)\hat{A}_1(t_0) \nonumber \\
& - i\epsilon\int_{t_0}^t dt^\prime  e^{-(i\omega_q+\epsilon^2\Gamma)(t-t^\prime)} (\hat{A}_2^\dagger(t^\prime) \hat{\xi}_1^f(t^\prime) - \hat{\xi}_2^{f\dagger}(t^\prime)\hat{A}_1(t^\prime)) \nonumber \\
&-\int_{t_0}^t dt^\prime  e^{-(i\omega_q+\epsilon^2\Gamma)(t-t^\prime)}\frac{i\epsilon}{\omega_q}\Big[2\varepsilon \hat{A}_2^\dagger(t^\prime)\hat{A}_1(t^\prime)\nonumber\\ 
&-2t_c(\hat{N}_1(t^\prime)-\hat{N}_2(t^\prime))\Big]\nonumber\left[\sum_{s=1}^{\infty}
\lambda_{s}^{ph} \left(\hat{B}_{s}^{ph\dagger}(t^\prime) + \hat{B}_{s}^{ph}(t^\prime)\right)\right] \nonumber \\
&- i\epsilon^2 g \int_{t_0}^t dt^\prime \Theta(t) e^{-(i\omega_q+\epsilon^2\Gamma)(t-t^\prime)} (\hat{N}_1(t^\prime)-\hat{N}_2(t^\prime)) \hat{b}_1(t^\prime)
\end{align}
Now, we note that the first four lines of above equation give the formally exact time evolution of $\hat{A}_2^\dagger\hat{A}_1$, when the DQD unit is not connected to the cavity unit, i.e, with $g_0=0$. Let us now define 
\begin{align}
\label{def_xiA}
&\hat{\xi}_A(t)= g\left(\hat{A}_2(t)^\dagger\hat{A}_1(t) \mid_{g_0=0}\right)\nonumber\\
&= g e^{i\hat{\mathcal{H}}^f (t-t_0)}\hat{A}_2(t_0)^\dagger\hat{A}_1(t_0)e^{-i\hat{\mathcal{H}}^f (t-t_0)},\nonumber\\
&\hat{\mathcal{H}}^f = \hat{\mathcal{H}}_{DQD} +\hat{\mathcal{H}}_{L} + \hat{\mathcal{H}}_{DQD-L}+\hat{\mathcal{H}}_{ph}+\hat{\mathcal{H}}_{DQD-ph}.
\end{align}
With this definition, Eq.(\ref{A2A1_exact_all}) becomes
\begin{align}
&\hat{A}_2^\dagger(t)\hat{A}_1(t)=\hat{\xi}_A(t)/g \nonumber \\
&- i\epsilon^2 g \int_{0}^t dt^\prime  e^{-(i\omega_q+\epsilon^2\Gamma)(t-t^\prime)} (\hat{N}_1(t^\prime)-\hat{N}_2(t^\prime)) \hat{b}_1(t^\prime).
\end{align}
Using above equation in equation of motion of $\hat{b}_1$ (Eq.(\ref{bosons_bath_integrated})), we have,
\begin{align}
&\frac{d\hat{b}_1}{dt} = -i\omega_0 \hat{b}_1 -i \lambda \hat{b}_2-i\epsilon^2 \hat{\xi}_A(t) -i \epsilon^2 \hat{\xi}_1^b(t) - \epsilon^4 \frac{\kappa_1}{2} \hat{b}_1 \nonumber \\
&+\epsilon^4 g^2 \int_0^t dt^\prime  e^{-(i\omega_q+\epsilon^2\Gamma)(t-t^\prime)} (\hat{N}_1(t^\prime)-\hat{N}_2(t^\prime)) \hat{b}_1(t^\prime).
\end{align}
It can be checked that $\hat{N}_\ell(t^\prime)=\hat{N}_\ell(t)+O(\epsilon^2)$, so that, up to $O(\epsilon^4)$, we have the following simplification in the second line of the above equation,
\begin{align}
&\epsilon^4 g^2 \int_0^t dt^\prime  e^{-i(\omega_q+\epsilon^2\Gamma)(t-t^\prime)} (\hat{N}_1(t^\prime)-\hat{N}_2(t^\prime)) \hat{b}_1(t^\prime) \nonumber \\
&\simeq \epsilon^4 g^2 \Delta\hat{N}(t) \sum_{p,\alpha=1}^2 \Phi_{\alpha \ell}^b \Phi_{\alpha p}^b \hat{b}_p(t)\int_0^t dt^\prime  e^{-(i\omega_q-i\omega_{\alpha}+\epsilon^2\Gamma)t^\prime},     
\end{align}
where $\Delta\hat{N}(t)\equiv \hat{N}_1(t)-\hat{N}_2(t)$, and again, we have used the definitions in Eq.(\ref{cavity_normal_modes}). Since the term is already $O(\epsilon^4)$, we can ignore the boson-fermion coupling in obtaining $\Delta\hat{N}(t)$. Up to $O(\epsilon^0)$, $\Delta\hat{N}(t)$ is a conserved quantity.  This means, assuming our initial state was a product of cavity and DQD units, in above equation, we can write 
\begin{align}
&\Delta\hat{N}(t)\simeq \langle\Delta\hat{N}(t)\rangle|_{g=0} \nonumber \\
&= Tr\left(\rho_{DQD}^{tot}e^{i\hat{\mathcal{H}}^f (t-t_0)}\Delta\hat{N}(t_0)e^{-i\hat{\mathcal{H}}^f (t-t_0)}\right). 
\end{align}
This can only be done if the DQD-cavity coupling is small. 

We are interested in time $(t-t_0)\gg 1/(\epsilon^2\Gamma)$. In this time regime, the DQD, when not connected to the cavity unit, will reach a non-equilibrium steady state (NESS). We thus have
\begin{align}
\label{b1_eqn}
&\frac{d\hat{b}_1}{dt} \simeq -i\omega_0 \hat{b}_1 -i \lambda \hat{b}_2-i\epsilon^2 \hat{\xi}_A(t) -i \epsilon^2 \hat{\xi}_1^b(t) - \epsilon^4 \frac{\kappa_1}{2} \hat{b}_1 \nonumber \\
&+\epsilon^4 g^2  \Delta N_{ss} \sum_{p,\alpha=1}^2  \frac{\Phi_{\alpha \ell}^b \Phi_{\alpha p}^b}{i(\omega_q-\omega_\alpha)+\epsilon^2\Gamma}\hat{b}_p(t).
\end{align}
where 
\begin{align}
\Delta N_{ss}= \lim_{t\rightarrow \infty} Tr\left(\rho_{DQD}^{tot} e^{i\hat{\mathcal{H}}^f (t-t_0)}\Delta\hat{N}(t_0) e^{-i\hat{\mathcal{H}}^f (t-t_0)} \right),
\end{align}
is the NESS expectation value of $\Delta\hat{N}(t)$ with $g=0$.
Thus, in Eq.(\ref{b1_eqn}), we have an effective gain coming from the DQD if $\Delta N_{ss}>0$. In other words, the DQD need to be population inverted. This is acheivable in the DQD set-up for the detuning $\varepsilon>0$ if conditions given in Eq.(\ref{conditions_energy_scales}) are satisfied, as can be separately checked. Associated with the gain is a fluctuation term, $\epsilon^2\hat{\xi}_A(t)$ coming from the DQD, thereby satisfying the fluctuation dissipation relation. We are interested in obtaining the dynamics of expectation values of cavity field operators and the cavity bilinears. For obtaining the expectation value of the cavity field operators, we need
\begin{align}
\langle\hat{\xi}_A(t)\rangle=g\langle \hat{A}_2^\dagger \hat{A}_1 \rangle_{ss},
\end{align}
where $\langle...\rangle_{ss}$ refers to NESS expectation value in absence of coupling to the cavity. For obtaining the cavity bilinears, we will require the following correlation function,
\begin{align}
\label{def_DQD_noise_corr}
\langle\hat{\xi}_A^\dagger(t_1)\hat{\xi}_A(t_2)\rangle = g^2\langle \hat{A}_1^\dagger(t_1)\hat{A}_2(t_1)\hat{A}_2^\dagger(t_2)\hat{A}_1(t_2)\rangle|_{g_0=0}.
\end{align}
To evaluate the above correlation function, we note the following result which can be derived using Eq.(\ref{A2A1_exact_all}), and integrating out the fermionic leads following exactly analogous procedure as used in the previous section for integrating out the bosonic baths, 
\begin{align}
\hat{A}_2^\dagger(t)\hat{A}_1(t)=    e^{-(i\omega_q+\epsilon^2\Gamma)(t-t^\prime)} \hat{A}_2^\dagger(t^\prime)\hat{A}_1(t^\prime) + O(\epsilon^2),
\end{align} 
with $t>t^\prime$. In above, we have kept the factor $\epsilon^2\Gamma$ in the exponent even though it is $O(\epsilon^2)$. This is because $(\epsilon^2\Gamma)^{-1}$ gives the time scale for the DQD to reach steady state, which in turn, gives the linewidth to the power spectrum of the noise appearing in the cavity due to the DQD. As we will see below, this leads to an $O(\epsilon^2)$ contribution in $\langle\hat{\xi}_A^\dagger(t_1)\hat{\xi}_A(t_2)\rangle$. The equivalent equation for $\hat{A}_1^\dagger(t)\hat{A}_2(t)$ is obtained by taking Hermitian conjugate. Now, to evaluate the correlation function in Eq.(\ref{def_DQD_noise_corr}), we consider two cases, $t_1<t_2$ and $t_1>t_2$ separately. If $t_1<t_2$, we use above formula to relate $\hat{A}_2^\dagger(t_2)\hat{A}_1(t_2)$ to $\hat{A}_2^\dagger(t_1)\hat{A}_1(t_1)$. If $t_1>t_2$, we use the Hermitian conjugate of above formula to related $\hat{A}_1^\dagger(t_1)\hat{A}_2(t_1)$ to $\hat{A}_1^\dagger(t_2)\hat{A}_2(t_2)$. For $t_1-t_0,t_2-t_0\gg 1/(\epsilon^2\Gamma)$, the combined result for both cases can then be written as
\begin{align}
\label{DQD_noise_correlations}
& \langle\hat{\xi}_A^\dagger(t)\hat{\xi}_A(t^\prime)\rangle\simeq \int_{-\infty}^\infty \frac{d\omega}{2\pi} P(\omega)   e^{i\omega(t-t^\prime)}, \\
& P(\omega)=g^2\langle \hat{N}_1 \rangle_{ss} \frac{2\epsilon^2\Gamma}{(\omega-\omega_q)^2+\epsilon^4\Gamma^2} \nonumber,
\end{align}
where we have used fermionic commutation relations for simplification and used the fact that for $V\gg \mu_1 \gg \frac{\omega_q}{2}$, the DQD cannot be doubly occupied, so $\langle \hat{N}_1\hat{N}_2 \rangle_{ss}=0$. In above, $P(\omega)$ is the power spectral density of the noise appearing in the cavity due to the presence of the DQD. Note that the leading contribution in the integral comes from $P(\omega_q)\propto \epsilon^{-2}$. Thus, $\epsilon^4\langle\hat{\xi}_A^\dagger(t)\hat{\xi}_A(t^\prime)\rangle\propto \epsilon^2$, as we previously said.

So, finally, after integrating out the bosonic baths, as well as the DQD unit, we have the following effective equation of motion of the cavity operators, 
\begin{align}
&\frac{d}{dt}\left(
\begin{array}{c}
\hat{b}_1\\
\hat{b}_2\\
\end{array}
\right) = -i\mathbf{H}_{eff}\left(
\begin{array}{c}
\hat{b}_1\\
\hat{b}_2\\
\end{array}
\right) -i \left(
\begin{array}{c}
\hat{\xi}^b_1(t)+\hat{\xi}_A(t)\\
\hat{\xi}^b_2(t)\\
\end{array}
\right),  \nonumber\\
& \mathbf{H}_{eff} = \mathbf{H}_{S}-i\mathbf{v^B}+i\mathbf{v^A} \nonumber \\
& \mathbf{v^B} =\left(\frac{\kappa_1+\kappa_2}{4}\right) \mathbb{I} + \left(\frac{\kappa_1-\kappa_2}{4}\right) \sigma_z,~~\mathbf{v^A} = \left(
\begin{array}{cc}
v_{11}^A & v_{12}^A\\
0 & 0\\
\end{array}
\right) \nonumber \\
& v_{11}^A = g^2 \Delta N_{ss} \sum_{\alpha=1}^2 \frac{({\Phi_{1\alpha}^b})^2}{i(\omega_q-\omega_\alpha)+\Gamma}~,\nonumber \\
&v_{12}^A = g^2 \Delta N_{ss} \sum_{\alpha=1}^2 \frac{\Phi_{1\alpha}^b\Phi_{2\alpha}^b}{i(\omega_q-\omega_\alpha)+\Gamma},  
\end{align}
where we have dropped the $\epsilon$'s for notational convenience. Above equation can be simplified to Eq.(\ref{def_H_eff}) of main text using Eq.(\ref{cavity_normal_modes}), the fact that $\omega_q=\omega_0$, $\lambda \ll \Gamma$ and neglecting $\hat{\xi}^b_\ell(t)$. We justify the reason for neglecting $\hat{\xi}^b_\ell(t)$ of the bosonic baths in following subsection.

\subsection{Reason for fluctuations of bosonic baths being negligible}\label{thermal_fluctuations}
The quantities $\hat{\xi}^b_\ell(t)$ embody the thermal and quantum fluctuations due to the presence of the bosonic baths. Since $\langle\hat{\xi}^b_\ell(t)\rangle=0$, they do not play any role in obtaining expectation values of cavity field operators. However, they do play a role in obtaining cavity bilinears via their two-time correlator (Eq.(\ref{bath_noise_correlations})). But, at low temperatures, we can write their two-time correlator as
\begin{align}
\langle \hat{\xi}_\ell^{b\dagger}(t)\hat{\xi}_m^b(t^\prime)\rangle\simeq \delta_{\ell m} \int_{0}^{1/\beta}\frac{d\omega}{2\pi}\mathfrak{J}_\ell (\omega)\mathfrak{n}^B(\omega)e^{i\omega(t-t^\prime)},
\end{align} 
where the upper limit of the intergral has been replaced by $1/\beta$ because the bose distribution suppresses the contribution from $\beta\omega\gg 1$. Since we assume that the frequency of the cavities $\omega_0$ satisfies $\beta \omega_0\gg 1$, this means the main contribution of the fluctuations come from bath frequencies which are extremely off-resonant with the cavity frequency. On the other hand, the DQD is in resonance with the cavity frequency, since $\omega_q=\omega_0$. As a result, the contribution from fluctuations of the gain medium is always much larger that that from the loss medium, i.e, the bosonic baths. This is why, for the choice of energy scales in our problem, we can neglect $\hat{\xi}^b_\ell(t)$. We have also numerically confirmed this by keeping the thermal fluctuations, and observing that they make no change in the results.


\section{Obtaining NESS results for DQD-unit}\label{DQD_RQME}


\subsection{The general Redfield Quantum Master Equation}
To obtain the NESS results for DQD-unit without the cavity-unit, we will take the approach of the Redfield Quantum Master Equation (RQME). For an arbitrary set-up of a system connected to a bath, the full set-up can be taken as isolated and described via the full system+bath Hamiltonian $\hat{\mathcal{H}}=\hat{\mathcal{H}}_S+\hat{\mathcal{H}}_{SB}+\hat{\mathcal{H}}_B$. Here $\hat{\mathcal{H}}_S$ is the system Hamiltonian, $\hat{\mathcal{H}}_B$ and $\hat{\mathcal{H}}_{SB}$ is the system-bath coupling Hamiltonian. The system-bath coupling can be written in the general form $\hat{\mathcal{H}}_{SB} = \epsilon \sum_{\ell} \hat{S}_\ell \hat{B}_\ell$, where $\hat{S}_\ell$ is some system operator and $\hat{B}_\ell$ is some bath operator and $\epsilon$ is a small parameter controlling the strength of system-bath coupling. The initial density matrix of the set-up $\rho_{tot}(0)$ is considered to be in product form $\rho_{tot}(0) = \rho(0) \otimes \rho_B$, where $\rho(0)$ is some initial state of the system, and $\rho_B$ is the initial state of the bath. The initial state of the bath is often taken to be the thermal state with respect to the bath Hamiltonian. The Redfield Quantum Master Equation (RQME) is obtained by writing down the equation of motion for the reduced density matrix of the system up to $O(\epsilon^2)$ under Born-Markov approx. If $\langle\hat{B}_\ell\rangle_B=0$, where $\langle...\rangle_B$ refers to the average taken only with respect to bath, is satisfied initially, then, the RQME is given by
\begin{align}
\label{RQME_general}
\frac{\partial\rho}{\partial t} &=i[\rho, \hat{\mathcal{H}}_S] \nonumber \\
&-\epsilon^2 \sum_{\ell,m} \int_0^\infty dt^{\prime} \Big\{ [\hat{S}_\ell,\hat{S}_m(-t^{\prime})\rho(t)]\langle \hat{B}_\ell\hat{B}_m(-t^\prime)\rangle_B  \nonumber \\ & +[\rho(t)\hat{S}_m(-t^{\prime}),\hat{S}_\ell]\langle \hat{B}_m(-t^\prime)\hat{B}_\ell\rangle_B \Big\}~,  
\end{align}
where $\hat{S}_m(t)=e^{i\hat{\mathcal{H}}_St} \hat{S}_m e^{-i\hat{\mathcal{H}}_S t}$, $\hat{B}_m(t)=e^{i\hat{\mathcal{H}}_Bt} \hat{B}_m e^{-i\hat{\mathcal{H}}_B t}$. This gives the leading order dissipative term. Note that if system-bath coupling Hamiltonian is $O(\epsilon)$, the dissipative part of RQME is $O(\epsilon^2)$. 

\subsection{The RQME for the DQD-unit}
We define the following quantities,
\begin{align}
&m=
\left(
\begin{array}{cc}
\cos(\theta) & -\sin(\theta)\\
-\sin(\theta) & -\cos(\theta)\\
\end{array}
\right),~\Phi=
\left(
\begin{array}{cc}
\cos(\theta/2) & \sin(\theta/2)\\
-\sin(\theta/2) & \cos(\theta/2)\\
\end{array}
\right), \nonumber \\
& F_{\alpha \nu}(\omega) = \sum_{\ell=1}^2 \frac{\Phi_{\alpha \ell} \Phi_{\nu \ell} \mathfrak{J}^f_{\ell}(\omega)\mathfrak{n}^F_\ell(\omega)}{2}, ~
\mathfrak{n}^F_{\ell}(\omega) = \frac{1}{e^{\beta(\omega-\mu_{\ell})}+1},\nonumber \\
& f_{\alpha \nu}(\omega) = \sum_{\ell=1}^2 \frac{\Phi_{\alpha \ell} \Phi_{\nu \ell}\mathfrak{J}^f_{\ell}(\omega)}{2} \nonumber \\
& F_{\alpha \nu}^{\Delta}(\omega) = \mathcal{P}\int \frac{d\omega^\prime}{\pi} \frac{F_{\alpha \nu}(\omega^\prime)}{\omega^\prime -\omega}, ~~ f_{\alpha \nu}^{\Delta}(\omega) = \mathcal{P}\int \frac{d\omega^\prime}{\pi} \frac{f_{\alpha \nu}(\omega^\prime)}{\omega^\prime -\omega},
\end{align}
where $\theta$ is as defined in Eq.(\ref{def_theta}).

The RQME for DQD-unit, when not connected to the cavity, is obtained by using Eq.(\ref{RQME_general}) and simplifying. The RQME is given by,
\begin{align}
&\frac{\partial \rho_{DQD}}{\partial t} = i[\rho_{DQD}, \hat{\mathcal{H}}_S] -\epsilon^2  \mathcal{L}_{ph} \rho_{DQD}  -\epsilon^2  \mathcal{L}_{f} \rho_{DQD},  \\
& \mathcal{L}_{ph} \rho_{DQD} \nonumber \\
&= \sum_{\alpha,\nu,\gamma,\delta} \left( m_{\alpha \nu} m_{\gamma \delta} [\hat{A}_\gamma^\dagger \hat{A}_\delta, \hat{A}_\alpha^\dagger \hat{A}_\nu \rho_{DQD}] F_B(\omega_\alpha^f-\omega_\nu^f) + h.c \right), \nonumber \\
& \mathcal{L}_{f} \rho_{DQD} = \sum_{\alpha,\nu} \left( [\hat{A}_\alpha^\dagger, \hat{G}_{\alpha\nu} \hat{A}_\nu \rho_{DQD}] + [\rho_{DQD} \hat{F}_{\alpha\nu}\hat{A}_\nu, \hat{A}_\alpha^\dagger] + h.c \right) \nonumber,
\end{align} 
where $\rho_{DQD}$ is the density matrix of the DQD, $h.c.$ refers to Hermitian conjugate, and 
\begin{align}
&\omega_1^f=\frac{\omega_q}{2}, ~~\omega_2^f=-\frac{\omega_q}{2}, \\
& F_B(\omega_\alpha^f - \omega_\nu^f) \nonumber\\
&= \int_0^\infty dt \int_0^\infty \frac{d\omega}{2\pi} e^{-i(\omega_\alpha^f - \omega_\nu^f)t}\mathfrak{J}_{ph}(\omega) \Big[\textrm{coth}(\frac{\beta\omega}{2})\cos(\omega t) \nonumber \\
&- i \sin(\omega t)\Big] \nonumber \\
& \hat{G}_{\alpha\nu} = \hat{f}_{\alpha\nu}-\hat{F}_{\alpha\nu}, \nonumber \\
& \hat{F}_{\alpha\nu} = \mathfrak{F}_{\alpha \nu} (\omega_\nu)(1-\hat{N}_{\bar{\nu}}) + \hat{N}_{\bar{\nu}} \mathfrak{F}_{\alpha \nu} (\omega_\nu + V), \nonumber \\
& \mathfrak{F}_{\alpha \nu}(\omega) =F_{\alpha \nu}(\omega)+i F_{\alpha \nu}^{\Delta}(\omega), \nonumber \\
& \hat{f}_{\alpha\nu} = \mathfrak{f}_{\alpha \nu} (\omega_\nu)(1-\hat{N}_{\bar{\nu}}) + \hat{N}_{\bar{\nu}} \mathfrak{f}_{\alpha \nu} (\omega_\nu + V) \nonumber \\
& \mathfrak{f}_{\alpha \nu}(\omega) =f_{\alpha \nu}(\omega)+i f_{\alpha \nu}^{\Delta}(\omega).
\end{align}
Here $\bar{\nu}$ is the complement of $\nu$ (i.e, if $\nu=1$, $\bar{\nu}=2$ and vice-versa, this convention will be followed throughout). We are interested in the regime, $V\gg\mu_1,\mu_2$. In this regime, there is negligible probability for double occupancy of the DQD, so $\langle \hat{N}_1 \hat{N}_2 \rangle\simeq 0$. Assuming this, and taking expectation values, we have
\begin{widetext}
\begin{align}
\label{N_eqn} \frac{d \langle\hat{N}_\gamma \rangle}{dt} =& - \sum_{\nu=1}^2 \Big[ m_{\gamma\bar{\gamma}}\Big(m_{\bar{\gamma}\nu} F_B(\omega^f_{\bar{\gamma}}-\omega^f_\nu) \langle \hat{A}_\gamma^\dagger \hat{A}_\nu \rangle  - m_{\gamma \nu} F_B(\omega^f_{\gamma}-\omega^f_\nu) \langle \hat{A}_{\bar{\gamma}}^\dagger \hat{A}_\nu \rangle +h.c. \Big) \Big]\nonumber \\
& -2 \Big [ \Gamma\langle \hat{N}_\gamma \rangle + F_{\gamma \gamma} (\omega^f_\gamma)(1-\hat{N}_{\bar{\gamma}}) \Big],~~~~~~~~\gamma=\{1,2\},\bar{\gamma}=\{2,1\}  \\
\label{A1A2_eqn}\frac{d \langle\hat{A}_1^\dagger \hat{A}_2 \rangle}{dt} =&-i\omega_q\langle\hat{A}_1^\dagger \hat{A}_2 \rangle- \sum_{\alpha,\nu=1}^2 \Big [ m_{2\alpha} \Big ( m_{\alpha \nu} F_B(\omega^f_\alpha - \omega^f_\nu) \langle\hat{A}_1^\dagger \hat{A}_\nu \rangle - m_{1\nu} F_B^*(\omega^f_1-\omega^f_\nu)\langle\hat{A}_\nu^\dagger \hat{A}_\alpha \rangle\Big) + (1 \leftrightarrow 2 )^\dagger \Big] \nonumber \\
&- \Big[\Gamma \langle \hat{A}_1^\dagger \hat{A}_2 \rangle  -  \mathfrak{F}_{21}(\omega^f_1)\Big(1-\langle \hat{N}_2 \rangle \Big)  + \mathfrak{F}_{12}^*(\omega^f_1)\Big(1-\langle \hat{N}_1 \rangle \Big)  \Big], 
\end{align}
\end{widetext}
where again, we have dropped the $\epsilon$'s for notational convenience.
To obtain $\langle\hat{N}_1 \rangle_{ss}$, $\langle\hat{N}_2 \rangle_{ss}$, $\langle\hat{A}_1^\dagger \hat{A}_2 \rangle_{ss}$, the above equations are solved with the LHS set to zero. $\langle\hat{A}_2^\dagger \hat{A}_1 \rangle_{ss}$ is complex conjugate of $\langle\hat{A}_1^\dagger \hat{A}_2 \rangle_{ss}$.

\section{General results for input-output experiment}\label{transmission_results}
Here we give the general expressions for the transmitted signal in either cavity in an input-output experiment with a coherent drive of frequency $\omega_d$ and strength $E$ at the second (lossy) cavity (see Eq.(\ref{eq:eom})). To this end, first we re-write the effective Hamiltonian (Eq.(\ref{def_H_eff})) in the rotating frame as
\begin{align}
\mathbf{H}_{\mathrm{eff}}=\left(\omega_0-\omega_d-\frac{i\kappa}{2}\right)\mathbbm{1}_2+\frac{\lambda}{2}(2-\delta)\sigma_x-i\frac{\lambda\delta}{2}\sigma_y+i\frac{\tilde\kappa}{2}\sigma_z, 
\end{align}
with the following definitions,
\begin{align}
\kappa=\frac{\kappa_2-\kappa^1_{eff}}{2},~\tilde{\kappa}=\frac{\kappa_2+\kappa^1_{eff}}{2},~\kappa_1^{eff}=2\Gamma\delta-\kappa_1.
\end{align}
The input-output experiment requires that the two coupled cavities are overall dissipative, so that they reach a steady state.
The transmitted signals $T_\ell$ at the two cavities are given by  
\begin{align}
&\left[
\begin{array}{c}
T_1/\kappa_1 \\
T_2/\kappa_2 \\
\end{array}
\right]
=i(\mathbf{H}_\mathrm{eff}-\omega_d\mathbbm{1}_2)^{-1}\left[
\begin{array}{c}
0 \\
1\\
\end{array}
\right].
\end{align}
The transmitted signal at $\ell$th cavity can be written as $T_\ell=|T_\ell|e^{i\phi_\ell}$, where $|T_\ell|$ is the transmission amplitude, and $\phi_\ell$ is the phase response. The explicit expression for the transmission amplitudes are given by
\begin{align}
&|T_1| =\frac{ \kappa_1 \lambda (1-\delta)  }{ \sqrt{ \left[\left(\omega_0-\omega_d\right)^2 - \left(\frac{\kappa}{2}\right)^2-\left(\Lambda_+-\Lambda_-\right)^2\right]^2 +\kappa^2(\omega_0-\omega_d)^2 }  }, \nonumber \\
& |T_2| =\frac{ \kappa_2 \left[(\omega_0-\omega_d)^2 + \left(\frac{\kappa_1^{eff}}{2}\right)^2 \right]   }{ \sqrt{ \left[\left(\omega_0-\omega_d\right)^2 - \left(\frac{\kappa}{2}\right)^2-\left(\Lambda_+-\Lambda_-\right)^2\right]^2 +\kappa^2(\omega_0-\omega_d)^2 }  }.
\end{align} 
The explicit expressions for the phase response are given by $\phi_1$ and $\phi_2$, which are extracted from the following equations,
\begin{align}
&\tan(\phi_1) = \frac{\kappa(\omega_0-\omega_d)}{ \left(\omega_0-\omega_d\right)^2 - \left(\frac{\kappa}{2}\right)^2-\left(\Lambda_+-\Lambda_-\right)^2 } \nonumber \\
& \tan(\phi_2)= \nonumber \\
& \frac{ \kappa_1^{eff}\left[\left(\omega_0-\omega_d\right)^2 - \left(\frac{\kappa}{2}\right)^2-\left(\Lambda_+-\Lambda_-\right)^2\right]+2\kappa(\omega_0-\omega_d)  }{ 2(\omega_0-\omega_d) \left[\left(\omega_0-\omega_d\right)^2 - \left(\frac{\kappa}{2}\right)^2-\left(\Lambda_+-\Lambda_-\right)^2 - \frac{\kappa_1^{eff}\kappa}{2}  \right] }.
\end{align}
The above results were used in deriving Eqs.(\ref{phi_2_individually_lossy}), (\ref{pi_by_2_condition_loss_induced}), (\ref{zero_condition_loss_induced}).

\section{Derivation of local Lindblad equation for the cavities}
\label{local Lindblad derivation}

In this section, we give the microscopic derivation of Eq.(\ref{LLQME1}. In order to derive this equation, we use Eq.(\ref{RQME_general}) treating the two cavities as the system, and the rest of the set-up as bath. This gives
\begin{align}
\label{QME1}
\frac{\partial\rho}{\partial t} &= i[\rho,\hat{\mathcal{H}}_C]-\epsilon^4\int_0^\infty dt \Big\{[\hat{b}_1, \hat{b}_1^\dagger(-t)\rho]\langle \hat{\xi}_A^\dagger(0) \hat{\xi}_A(-t) \rangle  \nonumber \\
&+\sum_{\ell=1,2} [\hat{b}_\ell, \hat{b}_\ell^\dagger(-t)\rho]\langle \hat{\xi}_\ell^{b^\dagger}(0) \hat{\xi}_\ell^b(-t) \rangle +{~\rm~h.c} \Big\},
\end{align}
where ${~\rm~h.c}$ refers to Hermitian conjugate, $\hat{b}_\ell(t)=e^{i\hat{\mathcal{H}}_Ct} \hat{b}_\ell e^{-i\hat{\mathcal{H}}_C t}$, $\hat{\mathcal{H}}_C$ is the Hamiltonian of the two coupled cavities, $\hat{\xi}_A(t)$ is defined in Eq.(\ref{def_xiA}), $\hat{\xi}^b_\ell(t)$ is defined in Eq.(\ref{def_xib}) and we have again explicitly put the small parameter $\epsilon$. Now, we assume that the coupling between the two cavities is also small,
\begin{align}
\lambda \rightarrow \epsilon \lambda.
\end{align}
so that, to $O(\epsilon^4)$, the hybridization between the two cavity modes can be ignored in the non-unitary part of the quantum master equation. So we use, 
\begin{align}
\hat{b}(t) \simeq e^{-i\omega_0 t}\hat{b} + O(\epsilon)
\end{align}
in Eq.(\ref{QME1}). This gives,
\begin{align}
\frac{\partial\rho}{\partial t} &= i[\rho,\hat{\mathcal{H}}_C]-
\int_0^\infty dt \Big\{ \int_{-\infty}^{\infty} \frac{d\omega}{2\pi} P(\omega)   e^{i(\omega-\omega_0)t}  [\hat{b}_1, \hat{b}_1^\dagger\rho] \nonumber\\
& +\sum_{\ell=1,2}\int_{0}^{\infty}\frac{d\omega}{2\pi}\mathfrak{J}_\ell (\omega)\mathfrak{n}^B(\omega)e^{i(\omega-\omega_0)t}[\hat{b}_\ell, \hat{b}_\ell^\dagger\rho]+{~\rm~h.c} \Big\},
\end{align}
where, again, for notational convenience we have dropped $\epsilon$, and have used Eqs.(\ref{bath_noise_correlations}) and (\ref{DQD_noise_correlations}).
Evaluating this equation along with the same approximation on bosonic bath spectral densities is discussed before Eq.(\ref{const_spec_approx}), gives Eq.(\ref{LLQME1}).

\bibliography{ref_PT_DQD}


\end{document}